\documentclass[11]{article}

\pdfoutput=1
	\usepackage[colorlinks,citecolor=blue,urlcolor=blue,linkcolor = blue]{hyperref}

	\usepackage[comma,sort&compress]{natbib}
	\usepackage{amssymb, amsmath,bm,graphicx}
	\usepackage{mathtools}
	\usepackage{setspace}
	\usepackage{mathrsfs,amsthm}
	\usepackage[normalem]{ulem}
	\usepackage{subcaption}
	\usepackage{times,bm}
	\usepackage{multirow}
	\usepackage{anyfontsize}
	\usepackage{graphicx,psfrag,epsf}
	\usepackage{float}
	\usepackage{tabularx}
	\usepackage{url}
	\usepackage{booktabs}
	\usepackage{hyperref}
	\usepackage{color}
	\usepackage{xcolor}
	\usepackage{caption}
	\usepackage{bm}
	\usepackage{bbm}
	\usepackage[mathscr]{euscript}
	\usepackage{bigstrut,enumerate}
	\usepackage[pagewise]{lineno}
	\usepackage{pdfpages}

	\renewcommand{\liminf}{\varliminf}
	\renewcommand{\limsup}{\varlimsup}
	
	\newtheorem{theorem}{Theorem}
	
	\newtheorem{lemma}{Lemma}
	\newtheorem{prop}{Proposition}	
	\newtheorem{proc}{Procedure}
	\newtheorem{remark}{Remark}
	\newtheorem{corollary}{Corollary}

	\newcommand{\I}{\ensuremath{{\text{I}}}}

	\newcommand{\thetab}{\ensuremath{{\bm{\theta}}}}

	\newcommand{\ex}{\ensuremath{\mathbb{E}}}
	\newcommand{\bs}[1]{\boldsymbol{#1}}
	\newcommand{\mbf}[1]{\mathbf{#1}} 
	 
	\newcommand{\mbb}[1]{\mathbb{#1}}
	\newcommand{\mfk}[1]{\mathcal{#1}}
	\newcommand{\mc}[1]{\mathcal{#1}}

	\newcommand{\prp}[1]{\textcolor{black}{#1}}
	
	\newcommand{\tbm}[1]{\textcolor{black}{#1}}
	\DeclareMathOperator*{\argmin}{arg\,min}

	\newcommand{\norm}[1]{\left\lVert#1\right\rVert}

\begin{document}
\setpagewiselinenumbers
%
%

\title{Adaptive Sparse Estimation with Side Information}
\author{Trambak Banerjee$^1$, Gourab Mukherjee$^1$ and Wenguang Sun$^{1, 2}$
	\\
	Department of Data Sciences and Operations, University of Southern California}

\date{}

\footnotetext[1]{The research of WS was supported in part by NSF grants DMS-CAREER 1255406 and DMS-1712983. {TB and GM were 
		partially supported by NSF DMS-1811866 and by the Zumberge individual award from the University of Southern California`s James H. Zumberge Faculty Research and Innovation Fund.}} 
\footnotetext[2]{Corresponding author: wenguans@marshall.usc.edu}
\maketitle

\begin{abstract}

The article considers the problem of estimating a high-dimensional sparse parameter in the presence of side information that encodes the sparsity structure. We develop a general framework that involves first using an auxiliary sequence to capture the side information, and then incorporating the auxiliary sequence in inference to reduce the estimation risk. The proposed method, which carries out adaptive SURE-thresholding using side information (ASUS), is shown to have robust performance and enjoy optimality properties. We develop new theories to characterize regimes in which ASUS far outperforms competitive shrinkage estimators, and  establish  precise conditions under which ASUS is asymptotically optimal. Simulation studies are conducted to show that ASUS substantially improves the performance of existing methods in many settings. The methodology is applied for analysis of data from single cell virology {studies} and microarray time course experiments. 

\end{abstract}
 
 \bigskip\bigskip


\newpage

\section{Introduction}
\label{sec1}

The recent technological advancements have made it possible to collect vast amounts of data with various types of side information such as domain knowledge, expert insights, covariates in the primary data, and secondary data from related studies. In a wide range of fields including genomics, neuroimaging and signal processing, incorporating side information promises to yield more accurate and meaningful results. However, few analytical tools are available for extracting and combining information from different data sources in high-dimensional data analysis. This article aims to develop new theory and methodology for leveraging side information to improve the efficiency in estimating a high-dimensional sparse parameter.   
We study the following closely related issues: (i) how to properly extract or construct an auxiliary sequence 
to capture useful sparsity information; (ii) how to combine the auxiliary sequence with the primary summary statistics to develop more efficient estimators; and (iii) how to assess the relevance and usefulness of the side information, as well as the robustness and optimality of the proposed method.  


\subsection{Motivating applications}

Sparsity is an essential phenomenon that arises frequently in modern scientific studies. In a range of data-intensive application fields such as genomics and neuroimaging, only a small fraction of data contain useful signals. The detection, estimation and testing of a high-dimensional sparse object have many important applications and have been extensively studied in the literature \citep{DonJin04, johnstone2004needles, Abretal06}. For instance, in the RNA-seq study that will be analyzed in Section \ref{analysis.sec}, the goal is to  estimate the true expression levels of $n=53,216$ genes for the virus strain VZV, which is the causative agent of varicella (chickenpox) and zoster (shingles) in humans \citep{zerboni2014molecular}. {The parameter of interest (the population mean vector of gene expression) is sparse as it is known that very few genes in the generic RNA-seq kits express themselves in these single-cell virology studies \citep{sen2018distinctive}.}
The accurate identification and estimation of nonzero large effects is helpful for the discovery of novel genetic biomarkers, which constitutes a key step in the development of new treatments and personalized medicine \citep{matsui2013genomic, holland2016estimating, erickson2005empirical}. Another example arises from microarray time-course (MTC) experiments that will be analyzed in Section E of the Supplementary Material. The goal is to identify genes that exhibit a specific pattern of differential expression over time. The temporal pattern, which can be revealed by estimating the differences in expression levels of genes between two time points, would help gain insights into the mechanisms of the underlying biological processes \citep{calvano2005network, sun2011multiple}. After baseline removal, the parameter of interest is the difference between two mean vectors that are both individually sparse. 

\begin{figure}[!h]
	\centering
	\includegraphics[width=1.02\linewidth]{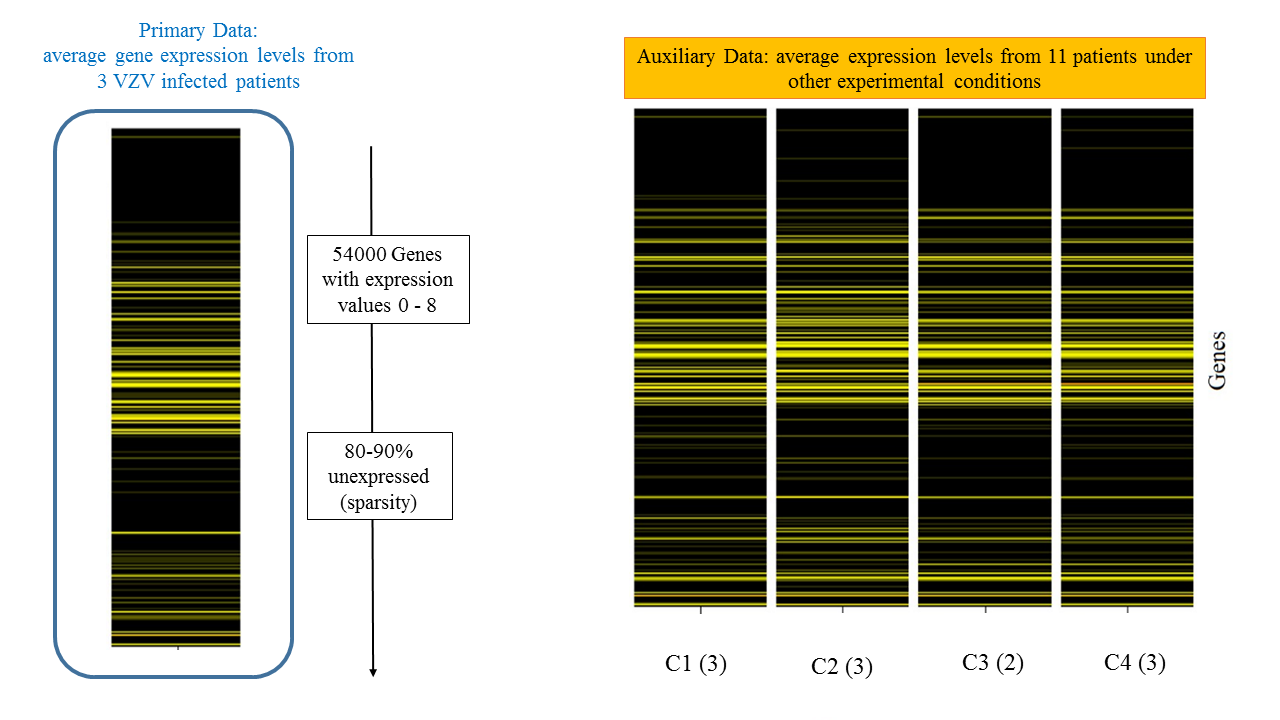}
	\caption{Heat map showing the average expression levels in the RNA-seq study. Left panel: VZV; right panel from top to bottom: 
	C1, C2, C3 and C4, where the number of replicates (patients) is shown in parenthesis. We can see that 80-90\% of the genes under the VZV condition are unexpressed (black), and the same sparse structure seems to be roughly maintained in the other four experimental conditions. Useful side information on sparsity can be extracted from secondary data (C1-C4) and be combined with the primary data (VZV) to construct more efficient estimators.}
	\label{VZV.fig}
\end{figure}

In practice, the intrinsic sparsity structure of the high-dimensional parameter is often captured by side information, which can be obtained as either summary statistics from secondary data sources or can be constructed as a covariate sequence from the original data. For instance, in the RNA-seq data, expression levels corresponding to other four experimental conditions (C1, C2, C3 and C4) are also available for the same $n$ genes through related studies conducted in the lab. The heat map in Figure \ref{VZV.fig} shows that the sparse structure of the mean transcription levels of the genes for VZV is roughly maintained by the same set of genes in subjects {from the other four conditions}. The common structural information shared by both cases (VZV) and controls (C1 to C4) can be exploited to construct more efficient estimation procedures. In the two-sample sparse estimation problem considered in the MTC study (analyzed in Section E of the Supplementary Material), we illustrate that a covariate sequence can be constructed from the original data matrix to assist inference by capturing the sparseness of the mean difference. Intuitively, incorporating side information promises to improve the efficiency of existing methods and interpretability of results. However, in conventional practice, such useful auxiliary data have been largely ignored in analysis. 

\bigskip 

\subsection{ASUS: a general framework for leveraging side information}

In this article, we develop a general integrative framework for sparse estimation that is capable of handling side information that may be extracted from (i) prior or domain-specific knowledge, (ii) covariate sequence based on the same (original) data, or (iii) summary statistics based on secondary data sources.
Let $\bm\theta=(\theta_1, \cdots, \theta_n)$ be an unknown high-dimensional sparse parameter.
Our study focuses on the class of non-linear thresholding  estimators [See Chs 8, 13 of \cite{johnstonebook} and  Ch 11 of \cite{Mallat-book}], which have been widely used in the sparse case where many coordinates of $\bm{\theta}$ are small or zero. 

The proposed estimation framework involves two steps: first constructing an auxiliary sequence $\bm S=(S_i: 1\leq i\leq n)$ to capture the sparse structure, and second combining $\bm S$ with the primary statistics, denoted $\pmb Y=(Y_i: 1\leq i\leq n)$, via a group-wise adaptive thresholding algorithm. Our idea is that the coordinates of $\pmb\theta$ become nonexchangeable in light of side information. To reflect this heterogeneity, we divide all coordinates into $K$ groups based on {$S_i$}. The side information is then incorporated in our estimation procedure by applying soft-thresholding estimators separately, thereby fine tuning the group-wise thresholds to capture the varied sparsity levels across groups. The optimal grouping and  thresholds are chosen adaptively via a data-driven approach, which employs the Stein's unbiased risk estimate (SURE) criterion to minimize the total estimation risk. The proposed method, which carries out adaptive SURE-thresholding using side information (ASUS), is shown to have robust performance and enjoy optimality properties. ASUS is simple and intuitive, but nevertheless provides a general framework for information pooling in sparse estimation problems. Concretely, since ASUS does not rely on any functional relationships between $\bm S$ and $\bm\theta$, it is robust and effective in leveraging side information in a wide range of scenarios. In Section \ref{principles.subsec}, we demonstrate that this flexible framework can be applied to various sparse estimation problems. 

The amount of efficiency gain of ASUS depends on two factors: (i) the usefulness of the side information; and (ii) the effectiveness in utilizing the side information. To understand the first issue, we formulate in Section \ref{theory.sec} a hierarchical model to assess the informativeness of an auxiliary sequence. Our theoretical analysis characterizes the conditions under which methods ignoring side information are suboptimal compared to an ``oracle'' with perfect knowledge on sparsity structure. To investigate the second issue, Section \ref{theory.sec} establishes precise conditions under which ASUS is asymptotically optimal, in the sense that its maximal risk is close to the theoretical limit that is attained by the oracle. Finally, we carry out a theoretical analysis on the robustness of ASUS; our results show that pooling non-informative side information would not harm the performance of data combination procedures. Our asymptotic results are built upon the elegant higher-order minimax risk evaluations developed by \citet{johnstone1994minimax}.

\subsection{Connections with existing work and our contributions}

ASUS is a non-linear shrinkage estimator that incorporates relevant side information by choosing data-adaptive thresholds to reflect the varied sparsity levels across groups. We use the SURE criterion for simultaneous tuning of the grouping and shrinkage parameters. Our methodology is related to \cite{xie2012sure}, \cite{tan2015improved} and \cite{weinstein2018group}, which utilized  SURE to devise algorithms reflecting optimal shrinkage directions. However, these works are developed for different purposes (addressing the heteroscedasticity issue in the data) and do not cover the sparse case.  

The notion of side information in estimation has been explored in several research fields. In information theory for instance, sparse source coding with side information is a well studied problem (\citealp{wyner1975source}; \citealp{cover2012elements}; \citealp{watanabe2015nonasymptotic}). However, these methodologies focus on very different goals and cannot be directly applied to our problem. In the statistical literature, the use of side information in sparse estimation problems has been mainly limited to regression settings where the side information must be in the form of a linear function of $\bm{\theta}$ \citep{ke2014covariance, kou2015optimal}. By contrast, our estimation framework utilizes a more flexible scheme that does not require the specification of any functional relationship between $\bm\theta$ and the side information. The proposed ASUS algorithm is simple and intuitive but nevertheless enjoys strong numerical and theoretical properties. Our simulation studies show that it can substantially outperform competitive methods in many settings. ASUS is a robust data combination procedure in the sense that asymptotically it would not under-perform methods ignoring side information when the auxiliary data are non-informative (see Theorem \ref{robust.thm}).
 
The proposed research makes several new theoretical contributions. First, we develop general principles for constructing and pooling the side information, which guarantees proper information extraction and robust performance of ASUS. Second, we formulate a theoretical framework to assess the usefulness of side information. Third, we establish precise conditions under which ASUS is asymptotically optimal. Finally, we extend the sparse minimax decision theory of \citet{johnstonebook}, which provides the foundation for a range of sparse inference problems  \citep{Abretal06, abramovich2007optimality,cai2014adaptive, tibshirani2014adaptive, collier2017minimax}, to derive new high-order characterizations of the maximal risk of soft-thresholding estimators.


\subsection{Organization of the paper}
Section \ref{method.sec} describes {the proposed ASUS procedure}. Section \ref{theory.sec} presents theoretical analyses. The numerical performances of ASUS are investigated using both simulated and real data in Section \ref{sec4}. Section \ref{sec:discuss} concludes with a discussion. Additional numerical results and proofs are given in the Supplementary materials.

\section{Adaptive Sparse Estimation with Side Information}
\label{method.sec}


This section first describes the model and assumptions (Section \ref{sec2.1}), then discusses how to construct the auxiliary sequence (Section \ref{principles.subsec}), and finally proposes the methodology (Section \ref{sec2.3}). 
 
\subsection{Model and assumptions}
\label{sec2.1}


To conduct a systematic study of the influence of side information for estimating $\bs{\theta}$, we consider a hierarchical model that relates the primary and auxiliary data sets through a latent vector $\bm{\xi}=(\xi_1,\ldots,\xi_n)$, which represents the noiseless side information that encodes the sparsity information of $\bs\theta$. The latent vector $\bm{\xi}$ cannot be observed directly but may be partially revealed by an auxiliary sequence (noisy side information) $\bm S=({S}_1, \cdots, {S}_n)$. 
For instance, in the RNA-seq example, the parameter of interest is the population mean of the gene expression levels for diseased patients, and the latent variable $\xi_i$ may represent the quantitative outcome of a complex gene regulation process that determines whether gene $i$ expresses itself under the influence of a certain experimental condition. The primary and 
{secondary} data respectively correspond to gene expression levels for the patients from {the 
concerned (i.e. VZV infected) and other related groups}. The primary and auxiliary statistics $Y_i$ and $S_i$ for gene $i$ can be constructed based on the corresponding sample means.   
 
%
For $n$ parallel units, the summary statistic $Y_i$ for the $i$th unit is modeled by
\begin{equation}
\label{y-theta}
Y_i=\theta_i+\epsilon_i, \quad \epsilon_i\sim N(0,\sigma_i^2),
\end{equation}
where, by convention, $\sigma_i^2$ are assumed to be known or can be well estimated from data (e.g. \citep{brown2009nonparametric, xie2012sure, weinstein2018group}). We further assume that both $\bm{\theta}$ and $\bm S$ are related to the latent vector $\bm{\xi}$ through some unknown real-valued functions $h_\theta$ and $h_s$:
\begin{eqnarray}
\label{theta-xi}
\theta_i & = & h_\theta(\xi_i, \eta_{1i}),
\\ S_i & = & h_s(\xi_i, \eta_{2i}),\label{s-xi}
\end{eqnarray}
where $\eta_{1i}$ and $\eta_{2i}$ follow some unspecified priors, and represent {independent} random perturbations that are independent of $\xi_i$; concrete examples for Models \ref{y-theta} to \ref{s-xi} are discussed in Section  \ref{principles.subsec}. 

\begin{remark}
	\label{rem1}
\rm{The above model can be conceptualized as a Bayesian hierarchical model: 
$$
Y_i|(\theta_i,S_i)\sim N(\theta_i,\sigma_i^2),\quad (\theta_i,S_i)|\xi_i\sim f_1(\theta|\xi_i)f_2(s|\xi_i),\quad \xi_{i}\stackrel{iid}{\sim}f_3(\xi),
$$ 
where $f_1,f_2,f_3$ are unknown densities. In Equations \ref{theta-xi} and \ref{s-xi}, $\xi_i$ is a random quantity and independent of $\eta_{1i}$ and $\eta_{2i}$. As a special case of Equation \ref{theta-xi}, we can write $\theta_i = h_\theta(\xi_i)$ without the random perturbations $\eta_{1i}$. Our theory is mainly stated in terms of random $\xi_i$'s for ease of presentation. However, we note that our theoretical results still hold even when $\xi_i$ is deterministic because the theory in Section \ref{sec2.3} is derived conditional on $\xi_i$, and the proof in Section \ref{theory.sec} is built upon an empirical density function \eqref{eq:3.4}. } 
\end{remark}

The hierarchical Models \ref{y-theta} to \ref{s-xi} provide a general and flexible framework for our methodological and theoretical developments. In particular, it covers a wide range of scenarios by allowing the strength of the side information to vary from completely non-informative (e.g., when $\xi_i$ is useless, or when $S_i$ and $\xi_i$ are independent for all $i$) to perfectly informative (e.g. when {$\theta_i=\xi_i$} and $S_i=\xi_i$ for all $i$). In Section \ref{theory.sec}, the usefulness of the latent vector $\bm \xi$ is  investigated via Equation \ref{theta-xi}, and the informativeness of the auxiliary sequence $\bm S$ is characterized by Equations \ref{theta-xi} and \ref{s-xi}. 



\subsection{Constructing the auxiliary sequence: principles and examples}
\label{principles.subsec}

A key step in our methodological development is to properly extract side information using an auxiliary sequence. The sequence $\pmb S$ can be constructed from various data sources including the following three basic settings: (i) prior or domain-specific knowledge; (ii) covariates or discard data in the same primary data set; or (iii) secondary data from related studies. We stress that our estimation framework is valid for all three settings as long as $\pmb S$ fulfills the following two fundamental principles. 

The first principle is \emph{informativeness}, which requires that $S_i$ should be chosen or constructed in a way to encode the sparse structure effectively. The second principle is \emph{conditional independence}, which requires that $S_i$ must be conditionally independent of $Y_i$ given the latent variable $\xi_i$. {The conditional independence assumption, which is implied by Models \ref{y-theta} to \ref{s-xi}, ensures proper shrinkage directions and plays a key role in establishing the robustness of ASUS.} 
Examples 1 to 4 below present specific instances of auxiliary sequences fulfilling such principles, wherein the auxiliary sequences may either be readily available from distinct but related experiments or can be carefully constructed from the same (original) data to capture important structural information that is discarded by conventional practice. 
 
\medskip

\noindent\emph{Example 1. Prioritized subset analysis (PSA, \citealp{li2008prioritized}).} In genome wide association studies, prior data and domain knowledge such as known gene functions or interactions may be used {to construct an auxiliary sequence $\bm S$ that can} 
prioritize the discovery of SNPs in certain genomic regions. Typically, the primary data set can be summarized as a vector $\bm Y=(Y_1, \cdots, Y_n)$, where $Y_i$ are either taken as differential allele frequencies between diseased and control groups, or z-values based on $\chi^2$-tests assessing the association between the allele frequency and the disease status. Let $\bm S=({S}_1, \cdots, {S}_n) \in \{-1, 1\}^n$ be an auxiliary sequence, where $S_i=1$ if SNP $i$ is in the prioritized subset and $S_i=-1$ otherwise. $\bm S$ can be viewed as perturbations of the true state sequence $\bm \xi=(\xi_1, \cdots, \xi_n)$, where $\xi_i=1$ if SNP $i$ is associated with the disease and $\xi_i=-1$ otherwise. The informativeness and independence principles are fulfilled when (i) the prioritized subset contains SNPs that are more likely to hold disease susceptible variants and (ii) the perturbations of $\bm \xi$ are random (hence $Y_i$ and $S_i$ are conditionally independent given $\xi_i$). Both (i) and (ii) seem reasonable assumptions in PSA studies. 

\medskip

\noindent\emph{Example 2. One-sample inference. } In the RNA-seq study, let {the primary data be} $\{Y_{i,j}: i=1, \cdots, n; j=1, \cdots, k_y\}$ that record the expression levels of $n$ genes from $k_y$ subjects infected by VZV. The primary statistics are $\bm Y=(\bar Y_1, \cdots, \bar Y_n)$, where $\bar Y_i=k_y^{-1}\sum_{j=1}^{k_y}Y_{i, j}$. Let {the secondary data be} $\{X_{i, j}: i=1, \cdots, n; j=1, \cdots, k_x\}$ {that} record the expression levels of the same $n$ genes for $k_x$ subjects {but under different} Conditions C1 to C4. The auxiliary sequence can be constructed as $\bm {S}=({S}_1, \cdots, {S}_n)=(|\bar X_1|, \cdots, |\bar X_n|)$, where $\bar X_i=k_x^{-1}\sum_{j=1}^{k_x} X_{i, j}$. {Thus although we record the expression levels of the same set of $n$ genes, in the case of the primary data the genes are infected with the VZV virus whereas for the secondary data the expression levels are recorded under the influence of agents that are different from that of the VZV virus.} The latent state $\xi_i$ represents whether gene $i$ expresses itself under any of the conditions. Now we check whether the two information extraction principles are fulfilled. First, the informativeness principle holds since, as demonstrated by the heat map in Figure \ref{VZV.fig},  inactive genes under VZV are likely to remain inactive under the other conditions. The sparse structure is captured by the auxiliary sequence, where a small $S_i$ signifies an inactive gene.  Second, {Section \ref{sec2.1} has explained how} the RNA-seq data may be sensibly conceptualized via Models \eqref{y-theta} to \eqref{s-xi}, where $\bar{Y}_i$ and $S_i$ are conditionally independent given the latent variable $\xi_i$, fulfilling the second principle.  

\medskip

\noindent\emph{Example 3. Two-sample inference. } Consider the MTC study discussed in the introduction (and analyzed in Section E of the Supplementary Material). Let $\{Y_{i,j, t_d}:i=1,\ldots,n;j=1,\ldots,k_i; d=0, 1, 2\}$ record the expression levels of $n$ genes from $k_i$ subjects at time points $t_0$ (baseline), $t_1$ and $t_2$. Let $\bar{Y}_{i, d}=k_i^{-1}\sum_{j=1}^{k_i}(Y_{i,j, t_d}-Y_{i,j,t_0})$ be the average expression levels of gene $i$ at time point $t_d$ after baseline adjustment, $d=1, 2$. Denote $\mu_{i,d}=E(\bar{Y}_{i, d})$ and $\pmb\mu_d=(\mu_{i,d}: 1\leq i\leq n)$.  
Then both $\pmb\mu_1$ and $\pmb\mu_2$ are individually sparse. The parameter of interest is $\theta_i=\mu_{i,1}-\mu_{i,2}$, which can be estimated by the primary statistic ${Y}_i=\bar{Y}_{i,1}-\bar{Y}_{i,2}$. Denote the union support $\mathcal U=\{i: \mu_{i,1}\neq 0 \; \mbox{or}\; \mu_{i,2}\neq 0\}$. Then $\mathcal U$ can be exploited to screen out zero effects since if $i\notin \mathcal U$, we must have $\theta_i=0$. 
Consider the sequence \tbm{$S_i=|\bar{Y}_{i, 1}+\kappa_i\bar{Y}_{i, 2}|$}, where $\kappa_i=\hat\sigma_{i,1}/\hat\sigma_{i,2}$ and $\hat\sigma_{i,d}^2=(k_i-1)^{-1}\sum_{j=1}^{k_i}(Y_{i,j, t_d}-Y_{i,j, t_0}-\bar{Y}_{i, d})^2$. Then the auxiliary sequence is informative since a large $S_i$ provides strong evidence that $i\in \mathcal U$. The union support encodes the sparse structure of $\pmb\theta$. Moreover, \tbm{$Y_i$ and $S_i$ are asymptotically independent with our choice of $\kappa_i$} (Proposition 6 in \citealp{cai2018cars}). Hence both principles are fulfilled.

\medskip

\noindent\emph{Example 4. Estimation under the ANOVA setting. } This example is an extension of Example 3 to multi-sample inference. Consider $m$ conditions $d=1,\ldots, m$, $m\geq 2$. The parameter of interest is $\bm{\theta}_{n\times 1}=\Gamma\mbf{a}$, where $\Gamma_{n\times m}=(\bm{\mu}_1,\dots,\bm{\mu}_m)$, $\mu_{i,d}= \mbb{E}(\bar{Y}_{i,d})$ and $\mbf{a}_{m\times 1}$ is a vector of known weights. Here $\bm{\theta}$ may represent a weighted average of true transcription levels of $n$ genes across $m$ time points. Let $\mbf{D}_i=(\bar{Y}_{i,1},\ldots,\bar{Y}_{i,m})$ be the vector of average expression level of gene $i$ for the $m$ time points after baseline adjustment and {denote $\mathbb{D}_{n\times m}=(\mbf{D}_1,\ldots,\mbf{D}_n)^{T}$. To estimate $\bm{\theta}$, our proposed framework suggests using the usual unbiased estimator $\pmb Y=\mbb{D}\mbf{a}$ as the primary statistic, and $\pmb S=\mbb{D}\mbf{b}$ as the auxiliary} 
{sequence} for some weights $\mbf{b}$. The informativeness principle from Example 3 continues to hold under this setting. To fulfil the independence principle, we choose $\mbf b$ such that \tbm{${Cov}(\pmb Y,\bm S)=\mbf{0}$}. 

\medskip

In Examples 3 and 4, the auxiliary sequence $\bm S$ is constructed from the same original data matrix. We give some intuitions to explain why $\bm S$ is useful. The conventional practice  reduces the original data into a vector of summary statistics $\bm{Y}$. However, this data reduction step often causes significant loss of information and thus leads to suboptimal procedures. Specifically, the information on the sparseness of the union support $\mathcal{U}$ is lost in the data reduction step. The key idea in Example 3 is that the auxiliary sequence $\pmb S$ captures the  structural information on sparsity, which is discarded by conventional practice. Therefore by incorporating $\pmb S$ into the inferential process we can improve the efficiency of existing methods. Note that $\bm Y$ is not a sufficient statistic for estimating $\pmb\theta$, the minimax estimation error based on $(\bm Y,\bm S)$ can greatly improve the performance of all estimators that are based on $\bm Y$ alone; a rigorous theoretical analysis is carried out in the proof of Theorem \ref{m.4}. To summarize, the above examples illustrate that the side information can be either ``external'' (Examples 1-2) or ``internal'' (Examples 3-4). The key in the proposed estimation framework, which we discuss next, is to construct a proper auxiliary sequence that fulfills the two fundamental principles. We shall develop a unified estimation framework that is capable of handling both internal and external side information. 

We conclude this section with two remarks. First, the conditional independence assumption can be relaxed; the methodology would work as long as $Y_i$ and $S_i$ are conditionally \emph{uncorrelated} (c.f. Proposition \ref{prop1}). Second, we do not require $Y_i$ or $\theta_i$ to be related to $S_i$ through any functional forms; hence classical regression techniques (even nonparametric models) cannot be applied in the above scenarios. We aim to develop a general information pooling strategy that does not involve any prescribed functional relationships; a methodology in this spirit is described next.

\subsection{The ASUS estimator and its risk properties}
\label{sec2.3}

Let $\pmb Y$ and $\bm S$ denote the primary statistics and auxiliary sequence obeying Models \eqref{y-theta} to \eqref{s-xi}. Let $\eta_{t}(.)$ be a soft-thresholding operator such that
	\begin{equation*}
	\eta_{t}(Y_i)=\begin{cases}
	-Y_i\sigma^{-1}_i\text{, if }|Y_i\sigma^{-1}_i|\le t;\\
	-t\text{ sign}(Y_i\sigma^{-1}_i)\text{, otherwise}.
	\end{cases}
	\end{equation*}
The proposed ASUS estimator operates in two steps: first constructing $K$ groups using $\bm S$, and second applying soft-thresholding within each group using $\bm Y$. The construction of the groups relies only on $\bm S$. The tuning parameters for both grouping and shrinkage are determined using the SURE criterion. 

\begin{proc} 
 For $k=1,\ldots,K$ and $\bm{\tau}=\{\tau_1<\ldots<\tau_{K-1}\}$, denote $\widehat{\mc{I}}_k^\tau=\{i:\tau_{k-1}<S_i\le\tau_{k}\}$ with $\tau_0=-\infty$, $\tau_K=\infty$. Consider the following class of shrinkage estimators:
\begin{equation}\label{ASUS-estimator}
\hat{\theta}^{SI}_{i}(\mfk{T})\coloneqq Y_i+\sigma_i \eta_{t_k}(Y_i)\text{ if }i\in\widehat{\mc{I}}_k^\tau,
\end{equation}
where, $\mfk{T}=\{\tau_1,\dots,\tau_{K-1},~t_1,\ldots,t_K\}$ and   
{each of the threshold hyper-parameters $t_1,\ldots,t_K$ varies in $[0,t_n]$ with $t_n=(2\log n)^{1/2}$. Thus, the set of all possible hyper-parameter $\mfk{T}$ values is $\mathcal{H}_n= \mbf{R}^{K-1}_+\times[0,t_n]^K$. Define the SURE function
\begin{equation}	\label{SURE:func}
	S(\mfk{T},\pmb Y,\bm S)=n^{-1}\left[\sum_{i=1}^{n}\sigma^2_i+\sum_{k=1}^{K}	\sum_{i\in \widehat{\mc{I}}_k^\tau} \left\{\sigma^2_i(|Y_i\sigma_i^{-1}|\wedge t_k)^{2}
	-2\sigma^2_i\,I(|Y_i\sigma_i^{-1}|\le t_k)\right\}\right].
\end{equation}
Let $\hat{\mfk{T}}=\argmin_{\mfk T \in \mathcal{H}_n}\,S(\mfk{T},\pmb Y,\bm S)$. Then, the ASUS estimator is given by $\hat{\theta}^{SI}_{i}(\hat{\mfk{T}})$.}
\end{proc}

\begin{figure}[!h]
	\centering
	\includegraphics[width=1.02\linewidth]{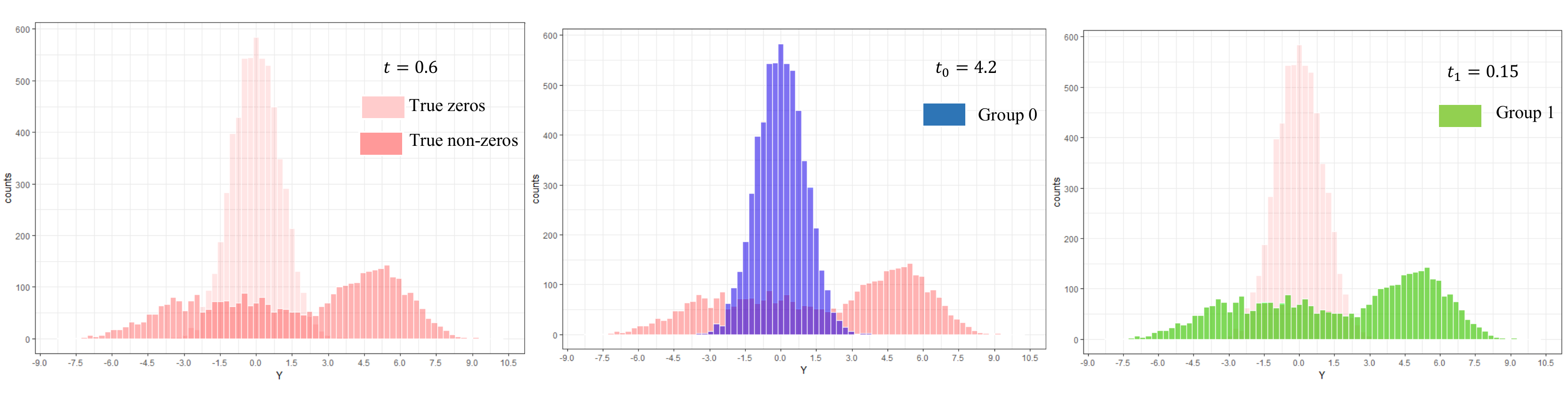}
	\caption{Toy example depicting ASUS. Left: SureShrink estimator at $t=0.6$. Middle: ASUS with group 0 and $t_0=4.2$. Right: ASUS with group 1 and $t_1=0.15$.}
	\label{fig3.1}
\end{figure}

\begin{remark}\rm{
When $\bm{\theta}$ is very sparse, the empirical fluctuations in the SURE function would have non-negligible effects on thresholding procedures. {We suggest choosing $t_1, \ldots, t_k$ for a given grouping by implementing a hybrid scheme that is similar to the SureShrink estimator of \cite{donoho1995adapting}, e.g. setting $t_k=t_n$ if $|\widehat{\mc{I}}_k^{\tau}|^{-1}\sum_{i\in \widehat{\mc{I}}_k^\tau}(Y_i^2/\sigma_i^2)\wedge t_n^2-1 \le n^{-1/2}\log^{3/2}n$.}}
\end{remark}
We present a toy example to illustrate why ASUS works. Consider the two-sample inference problem described by Example 3 in Section \ref{principles.subsec}. Let $\theta_i=\mu_{i,1}-\mu_{i,2}$ and $\bar{Y}_{i,d}\sim N(\mu_{i,d},0.25)$, where $d=1,2$, $i=1,\ldots, n$, and $n=10^4$. For $\bm{\mu}_{1}$ we generate the first $20\%$ of its coordinates randomly from $\mbox{Unif}(4,6)$, the next $20\%$ randomly from $\mbox{Unif}(2,3)$ and set the remaining coordinates to $0$. For $\bm{\mu}_{2}$, the first $20\%$ are from $\mbox{Unif}(1,2)$, the next $20\%$ from $\mbox{Unif}(1,6)$ and the remaining $0$. Finally, we let $\bar{Y}_i=\bar{Y}_{i,1}-\bar{Y}_{i,2}$  and $S_i= |\bar{Y}_{i,1}+\bar{Y}_{i,2}|$. The left panel in Figure \ref{fig3.1} presents the histogram of $\pmb Y =(\bar{Y}_i: 1\leq i\leq n)$, where the lighter shade corresponds to $\bar{Y}_i$ with $\theta_i=0$. The SureShrink estimator in \cite{donoho1995adapting} chooses threshold $t=0.6$ for all observations, resulting in an MSE of 0.338. Imagine that an oracle has the perfect knowledge about the two groups ($\theta_i=0$ vs. $\theta_i\neq 0$). In group 0, SureShrink chooses $t_0=4.2$, whereas in group 1, SureShrink chooses $t_0=0.15$. The total MSE is reduced to 0.20 by adopting varied thresholds for the two groups. In practice, the groups cannot be identified perfectly but can be partially revealed by the auxiliary statistic $S_i= |\bar{Y}_{i,1}+\bar{Y}_{i,2}|$, where a small $S_i$ signifies a possible zero effect. Our simulation studies in Section \ref{sec4} show that by exploiting the side information in $S_i$, ASUS achieves substantial gain in performance over conventional methods. 

Let {$l_n(\bm{\theta},\bm{\widehat{\theta}})=n^{-1}\|\bm{\widehat{\theta}}-\bm{\theta}\|^{2}_{2}$} denote the squared error loss of estimating $\bm{\theta}$ using $\widehat{\bm{\theta}}$. {For each member $\hat{\bm{\theta}}^{SI}(\mfk{T})$ in our class of estimators, $\mfk{T}\in\mathcal{H}_n$, denote its risk by} $r_n(\mfk{T};\bm{\theta})=\mbb{E}\left[l_n\left\{\bm{\theta},\hat{\bm{\theta}}^{SI}(\mfk{T})\right\}\right]$, where the expectation is taken with respect to the joint distribution of $(Y_i, S_i)$. The next proposition shows that  \eqref{SURE:func} provides an unbiased estimate of the true risk.
\begin{prop}
	\label{prop1}
Consider Models \eqref{y-theta} to \eqref{s-xi}. Then given $\xi_i$, the pair $\bigl\{(Y_i-\theta_i)\eta_{t_k}(Y_i),I(i\in\widehat{\mc{I}}_k^\tau)\bigr\}$ are uncorrelated. It follows that $r_n(\mfk{T};\bm{\theta})=\mbb E\{S(\mfk{T},\pmb Y, \bm S)\}$. 
\end{prop}
Next we study the large-sample behavior of the proposed SURE criterion. As in \citet{xie2012sure}, we impose the following assumption on the fourth moment of the noise distributions:
\begin{eqnarray*}
	\mathrm{(A1) }&&\limsup_{n\to\infty}\dfrac{1}{n}\sum_{i=1}^{n}\sigma_i^4<\infty~.
\end{eqnarray*}
 
{The following theorem shows that the risk estimate $S(\mfk{T},\pmb Y,\bm S)$ is uniformly close to the true risk as well as the loss, justifying our proposed hyper-parameter estimate $\mfk{\hat T}$. 
Compared to \citet{xie2012sure} (theorem 3.1) and \citet{brown2016empirical} (theorem 4.1), we obtain explicit rates of convergence by tracking the empirical fluctuations in the SURE function through sharper concentration inequalities.} 
 
\begin{theorem}
	\label{thm1}
	Under Assumption A1, with $c_n =n^{1/2}(\log n)^{-\delta}$ for any $\delta>3/2$, we have		
	\begin{align*}
	&\textrm{(a)}\quad \lim_{n \to \infty }\;\; c_n\;\;\mbb{E}\Big\{\sup_{\mfk{T}\in\mathcal{H}_n}\Big|S(\mfk{T},\pmb Y,\bm S)-r_n(\mfk{T};\bs{\theta})\Big|\Big\} = 0,\\
	&\textrm{(b)}\quad \lim_{n \to \infty }\;\; c_n\;\mbb{E}\Big[\sup_{\mfk{T}\in\mathcal{H}_n}\Big|S(\mfk{T},\pmb Y,\bm S)-l_n\{\bs{\theta},\hat{\bm{\theta}}^{SI}(\mfk{T})\}\Big|\Big]=0,
	\end{align*}
	where the expectation is with respect to the joint distribution of $\pmb Y,\bm S$.
	\end{theorem}

{Define $\mfk{T}^{OL}$ as the minimizer of the true loss function:
$\mfk{T}^{OL}=\argmin_{\mfk{T} \in \mathcal{H}_n} l_n\{\bs{\theta},\bs{\hat{\theta}}^{SI}(\mfk{T})\}$.
$\mfk{T}^{OL}$ is referred to as the \emph{oracle loss} hyper-parameter as it involves the knowledge of of $\bm{\theta}$. It provides the theoretical limit that one can reach if allowed to minimize the true loss. Let $\bs{\hat{\theta}}^{SI}(\mfk{T}^{OL})$ be the corresponding oracle loss estimator. The following corollary establishes the asymptotic optimality of $\hat{\mfk{T}}$.}

\begin{corollary}
	\label{cor1}
	Under assumption A1, if  $\lim_{n \to \infty}c_n\, {n^{-1/2}}\log^{\delta}n=0$ for any $\delta>3/2$, then\\
	\textrm{(a)} The loss of $\bs{\hat{\theta}}^{SI}(\hat{\mfk{T}})$ converges in probability to the loss of $\bs{\hat{\theta}}^{SI}(\mfk{T}^{OL})$: 
	\begin{equation*}
	\lim_{n\to\infty}\mbb{P}\left[l_n\left\{\bs{\theta},\bs{\hat{\theta}}^{SI}(\hat{\mfk{T}})\right\}\ge l_n\left\{\bs{\theta},\bs{\hat{\theta}}^{SI}(\mfk{T}^{OL})\right\}+c_n^{-1}\epsilon \right] =0\text{  for any }\epsilon>0~.
	\end{equation*}
	\textrm{(b)} The risk of $\bs{\hat{\theta}}^{SI}(\hat{\mfk{T}})$ converges to the risk of the oracle loss estimator:
	\begin{equation*}
	\lim_{n\to\infty} c_n\;\mbb{E}\left[l_n\left\{\bs{\theta},\bs{\hat{\theta}}^{SI}(\hat{\mfk{T}})\right\}- l_n\left\{\bs{\theta},\bs{\hat{\theta}}^{SI}(\mfk{T}^{OL})\right\}\right]=0~.
	\end{equation*}	
\end{corollary}

\subsection{Approximating the Bayes rule by ASUS}
\label{sec:empbayes}

This section discusses a Bayes setup and illustrates how ASUS may be conceptualized as an approximation to the Bayes oracle estimator.

Consider a hierarchical model where $\theta_i$ has an unspecified prior and $Y_i\stackrel{ind.}{\sim}N(\theta_i,\sigma^2_i)$ with $\sigma_i^2$ known. In the absence of any auxiliary sequence $\bm S$ and when $\sigma_i$ are all equal to, say $\sigma$, the optimal estimator is
\begin{equation}\label{opt-est}
\delta_i^\pi=E(\theta_i|y_i)=y_i+\sigma^2 \frac{f^\prime(y_i)}{f(y_i)},
\end{equation}
which is known as Tweedie's formula \citep{efron2011tweedie}. 
When the marginal densities $f(y_i)$ are unknown, \eqref{opt-est} can be implemented in an empirical Bayes (EB) framework. For example, \cite{brown2009nonparametric} used kernel methods to estimate unknown densities and showed that the resulting EB estimator is asymptotically optimal under mild conditions. Under the sparse setting, an effective approach to incorporate the sparsity structure is to consider, for example, spike-and-slab priors \citep{johnstone2004needles}. In decision theory it has been established that the posterior median is minimax optimal under spike-and-slab priors; see Thoerem 1 of \cite{johnstone2004needles}. Hence the soft-threshold estimators can be viewed as good surrogates to the Bayes rule under sparsity. When the sparsity level is unknown, the threshold should be chosen adaptively using a data-driven method. 

For a given pair of primary and auxiliary statistics $(Y_i, S_i)$, the Bayes oracle estimator is
\begin{equation}\label{opt-est2}
\delta_i^\pi=E(\theta_i|Y_i, S_i). 
\end{equation}
Conditionally on $S_i$, a Tweedie's formula for equation \eqref{opt-est2} can be written which would require estimating the conditional marginal densities $f(y_i|s_i)$ and its derivatives. ASUS can be viewed as a two-step approximation to the oracle estimator \eqref{opt-est2}. The first step involves using the auxiliary sequence to divide the $n$ coordinates into $K$ groups: $\delta_i^\pi \approx\hat\delta_k(Y_i) =E(\theta_i|Y_i, i\in G_k)=E(\theta_i|Y_i, S_i^*=k),$ which can be viewed as a discrete approximation to the oracle rule \eqref{opt-est2} by discretizing $S_i$ as a categorical variable $S_i^*$ taking values $k=1, \cdots, K$.  
The second step involves setting thresholds for separate groups to incorporate the updated structural information from the auxiliary sequence. This step makes sense because under the sparse regime, it is natural to use the class of soft-thresholding estimators as a convenient surrogate to the Bayes rule, and ideally the threshold should be set differently to reflect the varied sparsity levels across the groups. Finally the optimal grouping and optimal thresholds are chosen by minimizing a SURE criterion. 

This Bayesian interpretation reveals that ASUS may suffer from information loss in the discretization step. However, fully utilizing the auxiliary data by modeling $\bm S$ as a continuous variable is practically impossible under the ASUS framework since the search algorithm cannot deal with a diverging number of groups. Moreover, directly implementing \eqref{opt-est2} using bivariate Tweedie approaches is highly nontrivial and requires further research. ASUS, thus, seems to provide a simple, feasible yet effective framework to incorporate the side information.

\section{Theoretical Analysis}
\label{theory.sec}

This section studies the theoretical properties of ASUS under the important setting where $\bm \theta$ is sparse. By contrast, the results of Section \ref{sec2.3} hold for any sequence $\bm \theta$. To simplify the presentation, we focus on a class of thresholding estimators that utilize two groups. The two-group model provides a natural choice for some important applications such as the prioritized subset analysis and RNA-seq study, but the proposed ASUS framework can handle more groups. The major goal of our theoretical analysis is to gain insights on sparse inference with side information, for which the simple two-group setup helps in two ways. First, it leads to a concise and intuitive characterization of the potential influence of side information on simultaneous estimation. Second, it enables us to develop precise conditions under which ASUS is asymptotically optimal. 

\subsection{Asymptotic set-up}
\label{sec3.1}

Consider hierarchical Models \eqref{y-theta} to \eqref{s-xi}.  
We begin by considering an oracle estimator $\tilde{\bm{\theta}}_n^{SI}(\mfk{T}_n^{OR})$ that directly uses the noiseless side information  $\bm{\xi}$: 
\begin{equation}
\label{eq:3.2}
\tilde{\theta}^{SI}_{i,n}(\mfk{T}^{OR}_n)\coloneqq\begin{cases}
Y_i+\sigma_i \eta_{t_1^*}(Y_i)\text{ if }i\in\mc{I}_{1,n}^{\tau^\star},\\
Y_i+\sigma_i \eta_{t_2^*}(Y_i)\text{ if }i\in\mc{I}_{2,n}^{\tau^\star},
\end{cases}
\end{equation}
where $\mc{I}_{1,n}^{\tau}=\{i:\xi_i\le \tau\}$, $\mc{I}_{2,n}^{\tau}=\{i:\xi_i> \tau\}$, and
\begin{equation}
\label{eq:3.3}
\mfk{T}^{OR}_n:=(\tau^\star_n,t_{1,n}^*,t_{2,n}^*)=\argmin_{\mfk{T}\in\mbf{R}\times[0,t_n]\times[0,t_n]}\,\mbb{E} \,l_n\left\{\bm{\theta},\tilde{\bm{\theta}}^{SI}(\mfk{T})\right\}.
\end{equation}

\begin{remark}\rm
{Both the oracle estimator $\tilde{\bm{\theta}}_n^{SI}(\mfk{T}_n^{OR})$ and the oracle loss estimator $\bs{\hat{\theta}}^{SI}(\mfk{T}^{OL})$ assume the knowledge of $\pmb\theta$. However, they are different in that the former creates groups based on $\bm{\xi}$, whereas the latter uses $\pmb S$. The purposes of introducing these two oracle estimators are different: $\bs{\hat{\theta}}^{SI}(\mfk{T}^{OL})$ is used to assess the effectiveness of the SURE criterion; by contrast, $\tilde{\bm{\theta}}_n^{SI}(\mfk{T}_n^{OR})$ is employed to evaluate the usefulness of the noiseless side information, i.e. the maximal improvement in performance that can be achieved by incorporating $\pmb\xi$.}
\end{remark}

Denote $\pi_{1,n}=n^{-1}\sum_{i=1}^n\mc{I}(\xi_i\le \tau^\star_n)$ and $\pi_{2,n}=1-\pi_{1,n}$. Intuitively, the optimal partition $\tau_n^\star$ (within the class of thresholding procedures utilizing two groups) is chosen to maximize the ``discrepancy'' between the two groups.  For units in group $\mc{I}_{k,n}^{\tau^\star}$, the mixture density of $\theta_i$ is given by
\begin{eqnarray}
\label{eq:3.4}
g_{k,n}(\theta)=(1-p_{k,n})\,\delta_0\,+\,p_{k,n}\,{h_{k,n}}(\theta),  \quad k=1,2, \label{eq:the4}
\end{eqnarray}
where $\delta_0$ is a dirac delta function (null effects), $h_{k,n}$ is the (alternative) {empirical} density of non-null effects. Following remark \ref{rem1}, our theory developed based on the empirical density \eqref{eq:3.4} can handle both random and deterministic models; this can be more clearly seen in our proofs of the theorems. {Here} $p_{k,n}$ is the conditional proportion of non-null effects for a given group {and}  may be conceptualized as the probability that a randomly selected unit in group $\mc{I}_{k,n}^{\tau^\star}$ is a non-null effect. 

We consider an asymptotic set-up based on the sparse estimation framework in chapter 8.6 of \citet{johnstonebook}, which has been widely used in high-dimensional sparse inference \citep{johnstone1997wavelet, Abretal06, donoho1998minimax, mukherjee2015exact, tony2017optimal}. 
Let $p_{1,n}=n^{-\alpha}$ and $p_{2,n}=n^{-\beta}$ for some $0<\alpha<\beta\leq 1$. Define
 $\rho_n=\pi^{-1}_{1,n}\pi_{2,n}$. Consider the following parameter space $${\Theta_n(\alpha,\beta,\rho_n)}=\big\{\bm{\theta}\in\mbb{R}^n:\norm{\bm{\theta}}_{0}\le {n(n^{-\alpha}+\rho_n n^{-\beta})}/{(1+\rho_n)}\big\}.$$ The maximal risk of ASUS over $\Theta_n(\alpha,\beta,\rho_n)$ is 
 $$
 \mc{R}^{AS}_n(\alpha,\beta,\rho_n)=\sup_{\thetab \in \Theta_n(\alpha,\beta,\rho_n) } r_n( \mfk{\hat T},{\bm{\theta}}).
 $$  
Correspondingly, over the same parameter space $\Theta_n(\alpha,\beta,\rho_n)$, we let $\mc{R}^{OS}_n(\alpha,\beta,\rho_n)$ denote the maximal risk of the oracle procedure $\tilde {\bm{\theta}}^{SI}_n(\mfk{T}^{OR}_n)$, and $\mc{R}^{NS}_n(\alpha,\beta,\rho_n)$ the minimax risk of all soft thresholding estimators without side information. 
 
The risk difference $\mc{R}^{NS}_n-\mc{R}^{OS}_n$ is a key quantity that will be used in later analysis as the benchmark decision theoretic improvement due to incorporation of side information. Specifically, the noiseless side information $\pmb\xi$ is \emph{useful} if it provides non-negligible improvement on the risk: 
\begin{equation}\label{useful-xi}
\lim_{n \to \infty} n(\mc{R}^{NS}_n-\mc{R}^{OS}_n)=\infty.
\end{equation}
Moreover, the ASUS estimator is \emph{asymptotically optimal} if its risk improvement over $\mc{R}^{NS}_n(\alpha,\beta,\rho_n)$ is asymptotically equal to that of the oracle: 
\begin{equation}\label{optimal-asus}
\mathcal{RI}_n=\frac{\mc{R}_n^{NS}-\mc{R}_n^{AS}}{\mc{R}_n^{NS}-\mc{R}_n^{OS}}\rightarrow 1 \mbox{ as $n \rightarrow \infty$}. 
\end{equation}

\subsection{Usefulness of side information} 

We focus on Model \eqref{eq:3.4}, a hypothetical model based on the oracle partition $\tau_n^\star$. We state a few conditions that are needed in later analysis; some are essential for characterizing the situations where the side information is useful, i.e.  the oracle estimator $\tilde{\bm{\theta}}_n^{SI}(\mfk{T}_n^{OR})$ would provide non-negligible efficiency gain over competitive estimators.  

\begin{description} 
 \item (A2.1) $\lim_{n\to\infty}\rho_n \, n^{-\gamma_0}=0$ for some $\gamma_0 < \beta - \alpha$. 
 
 \item (A2.2)  For some $\nu < 1/2$ and $k_n=\log n$, $\lim_{n \to \infty} k_n^{\nu} (1-\pi_{1,n})=\infty$.
 
\item (A2.3) For some $\nu < 1/2$, $\lim_{n \to \infty} n^{\nu}\, \pi_{1,n} p_{1,n} \ = \infty$.  
 
 \item (A2.4)  Let $\bar{\sigma^2_n} = n^{-1}\sum_{i=1}^n \sigma_i^2$ and $ 0 < \liminf_{n \to \infty } \bar{\sigma^2_n} \leq \limsup_{n \to \infty } \bar{\sigma^2_n} < \infty$. 
 
\end{description}

\begin{remark}\rm{
(A2.1) implies $\pi_{2,n}\,p_{2,n}/(\pi_{1,n}\,p_{1,n}) \to 0$, {which ensures that the oracle partition is effective in the sense that the two resulting groups have different  sparsity levels.} 
The asymmetric condition can be easily flipped for generalization. (A2.2) is a mild condition which allows $\pi_{1,n}$ to approach $1$ but at a controlled rate. (A2.3) prevents the trivial setting where ASUS reduces to the SureShrink procedure with universal threshold $\sqrt{2 \log n}$, i.e. the side information would not have any influence in the estimation process. See lemma 3 (section B supplementary material) which shows that if $\varlimsup_{n\to\infty}~n^{1/2}\pi_{1,n}p_{1,n}<\infty$, then ASUS reduces to the SureShrink procedure, i.e. there is no need for creating groups. (A2.4) is a mild condition that is satisfied in most real life applications. }
\end{remark}
 
Now we study the usefulness of the noiseless side information. Following the theory in \citet{johnstone1994minimax}, the next theorem explicitly evaluates the risk difference $\mc{R}^{NS}_n-\mc{R}^{OS}_n$ up to higher order terms. The analysis overcomes the crudeness of the first order asymptotics {for evaluating thresholding rules} as pointed out by \cite{bickel1983minimax} and \cite{johnstone1994minimax}. 
\begin{theorem}\label{m.4}
Consider  the oracle estimator defined in \eqref{eq:3.2}-\eqref{eq:3.3}.  Under assumption A2.1, with  $k_n= \log n$, for all $\nu < 1$, we have, 
$$	\mc{R}^{NS}_n(\alpha,\beta,\rho_n)-\mc{R}^{OS}_n(\alpha,\beta,\rho_n)=\pi_{1,n}\, p_{1,n}\, \bar{\sigma^2_n} \left\{ \log \pi_{1,n}^{-1}(2 - 3 \alpha^{-1}k_n^{-1}) + O(k_n^{-\nu})\right\}.
$$
It follows from (A2.3) that $\lim_{n \to \infty} n(\mc{R}^{NS}_n-\mc{R}^{OS}_n)=\infty$, establishing  \eqref{useful-xi}.
\end{theorem}

\subsection{Asymptotic optimality of ASUS}

To evaluate the efficiency of ASUS, we need to compare the segmentation used by ASUS with that used by the oracle estimator. For a given segmentation hyper-parameter $\tau$, define 
$$\tilde q^{jk}_{i,n}(\tau)\coloneqq
\mbb{P}_n(\hat I_i^j|I_i^k\Big) \text{ for } j, k\in\{1,2\}, \;i = 1, \ldots,n,
$$
where $\hat I_i^1=\{S_i\le {\tau}\}$, $I_i^1=\{\xi_i\le \tau^\star_n\}$, $\hat I_i^2=\mbf{R}\setminus \hat I_i^1$, $\I_i^2=\mbf{R}\setminus I_i^1$, and the probability operator $\mbb{P}_n$ is based on Model \eqref{eq:3.4}.  Let
$$ q^{jk}_{i,n}(\tau)=\tilde q^{jk}_{i,n}(\tau) \quad \text{ if } \inf_{\tau \in \mbf{R}} \pi_{2,n} \tilde q_n^{12}(\tau)+\pi_{1,n}\tilde q_n^{21}(\tau) <  \inf_{\tau \in \mbf{R}} \pi_{1,n} \tilde q_n^{11}(\tau)+\pi_{2,n}\tilde q_n^{22}(\tau)$$ and otherwise
$ q^{jk}_{i,n}(\tau)=\tilde q^{kk}_{i,n}(\tau)$ and $q^{kk}_{i,n}(\tau)=1-q^{jk}_{i,n}(\tau)$ for $j\neq k$. Denote the weighted average
$$
q^{jk}_n(\tau)=\frac{\sum_{i=1}^n q^{jk}_{i,n}(\tau) \sigma_i^2} {\sum_{i=1}^n \sigma_i^2}, \quad j, k\in\{1,2\}.
$$

Viewing the data-driven grouping step of ASUS as a classification procedure with the oracle segmentation corresponding to the true states, we can conceptualize $q^{21}_n(\tau_n)$ and $q^{12}_n(\tau_n)$ as misclassification rates. Define the efficiency ratio 
\begin{equation}
\mathcal{E}_n=\frac{\mc{R}_n^{NS}-\mc{R}_n^{OS}}{\mc{R}_n^{AS}-\mc{R}_n^{OS}}.
\end{equation}
{For notational simplicity, the dependence of this ratio on $\alpha, \beta, \rho_n$ is not explicitly marked.} 
It follows from \eqref{optimal-asus} that $\mc{RI}_n=1-\mc{E}_n^{-1}$. Hence a larger $\mc{E}_n$ signifies better performance of ASUS. In particular, $\mc{E}_n \to \infty$ implies the asymptotic optimality of ASUS. The poly-log rates in the following theorem are sharp. 

\begin{theorem}
	\label{m.1} Assume (A2.1) -- (A2.4) hold. Let $k_n=\log n$. If there exists a sequence $\{\tau_n\}_{n\geq 1}$ such that 
\begin{equation}\label{mis-class}	
	\lim_{n \to \infty} k_n^{2} \, q^{21}_n(\tau_n)=0  \text{ and }  \lim_{n \to \infty}\rho_n\, q^{12}_n(\tau_n)=0,
\end{equation}	
	then ASUS is asymptotically optimal. In particular, for all $\nu<1$ we have		 	
\begin{equation}\label{en}
\liminf_{n \to \infty} k_n^{-\nu} \mc{E}_n \ge\,  2 \liminf_{n \to \infty}\log\pi_{1,n}^{-1}.
\end{equation} 
\end{theorem}

Next we present two hierarchical models, respectively with sub-Gaussian (SG) and sub-Exponential (SExp) tails, under which the misclassification rates can be adequately controlled. Let  $S_i|\xi_i$ be independent random variables with $\mu_i:=\mu_i(\xi_i)$ and $(\nu_i(\xi_i),b_i(\xi_i))$ such that 
$\ex\, \{\exp(\lambda(S_i-\mu_i))\}\leq \exp(\nu_i^2\lambda^2/2) \text{ for all i and all } |\lambda| \leq b_i^{-1}.$
Let $\limsup_{i} b_i < \infty$, $\limsup_{i} \nu_i < \infty$ and $\bar{b}_n=\sup_{1\leq i \leq n} \max(2 \nu_i^2,b_i)$. 
When $b_i=0$, the distribution of ${S}_i$ has sub-Gaussian tails.  
For two partitions $A$ and $B$ of the set $\{1,\ldots,n\}$, define the $\ell_1$ distance between the two sets 
$\{\mu_i: i \in A\}$ and $\{\mu_i: i \in B\}$ by $\textrm{dist}(A,B)=\inf\{|x-y|: x \in A, y \in B\}$.
Let $c_n= \bar{b}_n (2\log k_n+ \log \rho_n)$. The following lemma provides a sufficient condition under which the requirements on misclassification rates \eqref{mis-class} are satisfied. The proof of the lemma follows directly from the standard bounds for sub-Gaussian and sub-Exponential tails. 



\begin{lemma}
	\label{m.3}
Let $I^*_{1,n}=\{i:\xi_i\leq\tau^\star_n\}$ and $I^*_{2,n}=\{1,\ldots,n\}\setminus I^*_{1,n}$. The requirements on misclassification rates given by \eqref{mis-class} are satisfied if $$\liminf_{n \to \infty } c_n^{-\gamma} \textrm{dist}(I^*_{1,n}, I^*_{2,n}) > \gamma, $$ where $\gamma$ is $1/2$ if $\sup_{i} b_i = 0$ and $1$ otherwise. 
\end{lemma}

\subsection{Robustness of ASUS}
 
This section carries out a theoretical analysis to address the concern whether the performance of data combination procedures would deteriorate when pooling non-informative auxiliary data. We first characterize asymptotic regimes under which auxiliary data are non-informative (while the attention is confined to the prescribed class of two-group ASUS estimators), and then show that under such regimes, ASUS is robust in performance in the sense that it does not under-perform standard soft-thresholding methods.

\begin{theorem}\label{robust.thm}
		Suppose (A2.1) -- (A2.4) hold. Let $\rho_n=n^{\gamma_0}$ and $k_n=\log n$. 	
\begin{description}		
		\item (a)  Consider the following situations: (i) $\lim_{n \to \infty} k_n^{-1} \rho_n q^{21}_n(\tau_n)=\infty$; and (ii) $\lim_{n \to \infty} n \rho_n q^{21}_n(\tau_n)  =0$ but $\lim_{n \to \infty} k_n^{-1} \rho_n \,q^{12}_n(\tau_n)  = \infty$. If for all sequence $\{\tau_n\}_{n \geq 1}$ either (i) or (ii) holds, then we must have $\lim_{n\to\infty} \mc{E}_n = 1.$ Hence, the auxiliary data are non-informative.
	   \item  (b) We always have $\liminf \limits_{n\to\infty} \mc{E}_n \geq 1.$ {Thus, even when pooling non-informative auxiliary data ASUS would be at least as efficient as competing soft thresholding based methods that do not use auxiliary data.}  
\end{description}
\end{theorem}


{Our next result characterizes the performance of soft-thresholding estimators, where their efficacies are measured by the ratio of their respective maximal risks with respect to that of the oracle. The subsequent analysis is carried out using the ratios $\mc{R}^{AS}_n\big/\mc{R}_n^{OS}$ and $\mc{R}^{NS}_n\big/\mc{R}_n^{OS}$, instead of the ratios of the risk differences (e.g. $\mc{RI}_n$ and $\mc{E}_n$). In this metric, we see that any optimally tuned soft-thresholding procedure is robust; but the improvement due to the incorporation of the side information can be observed in the varied convergence rates.} 
Concretely, we show that the maximal risk of any soft thresholding scheme lies within a constant multiple of the oracle risk $\mc{R}^{OS}_n$ irrespective of the informativeness of the side information. Particularly, if $\liminf_{n \to \infty} \pi_{1,n}>0$, then $\lim_{n \to \infty} k_n^{\nu}\,({\mc{R}^{NS}_n}\big/{\mc{R}_n^{OS}}- 1) = 0$ for all $\nu <1$. By contrast, ${\mc{R}^{AS}_n}\big/{\mc{R}_n^{OS}}$ tends to $1$ at a  faster rate under the conditions of Theorem \ref{m.1}.  
%

\begin{lemma}
	\label{m.1a}
Let $c_n=\log \pi_{1,n}^{-1}/\{\alpha k_n-1.5\log (2\alpha k_n)+2.5+\log \phi(0)\}$ and $k_n=\log n$.
For any $\nu < 1$, under assumptions (A2.1) -- (A2.4), we have 
\begin{eqnarray*}	
	\lim_{n \to \infty} k_n^{2 \nu}\,\bigl\{\,{\mc{R}^{NS}_n}\big/{\mc{R}_n^{OS}}- \min(1+c_n,\;\beta/\alpha)\,\bigr\} & = & 0; \\
\limsup_{n \to \infty} k_n^{2 \nu}\,\bigl\{\,{\mc{R}^{AS}_n}\big/{\mc{R}_n^{OS}}- \min(1+c_n,\;\beta/\alpha)\,\bigr\} & \leq & 0.
\end{eqnarray*}
Under the conditions of Theorem \ref{m.1}, if there exists $\delta >0$ such that $\lim_{n \to \infty} k_n^{\delta} \log \pi_{1,n}^{-1} = \infty$, then 
	$$
	\lim_{n \to \infty} k_n^{ 1+\delta}\,({\mc{R}^{NS}_n}\big/{\mc{R}_n^{OS}}- 1) = \infty \quad \mbox{ and} \quad \lim_{n \to \infty} k_n^{2 \nu}\,({\mc{R}^{AS}_n}\big/{\mc{R}_n^{OS}}- 1) = 0.
	$$
Hence the risk of ASUS approaches the oracle risk at a faster rate.  	
\end{lemma}


\section{Numerical Results}
\label{sec4}
%
In this section we compare the performance of ASUS against several competing methods, including (i) the SureShrink (SS) estimator in Donoho and Johnstone (1995), (ii) the extended James Stein estimator (EJS) discussed in \citet{brown2008season}, (iii) the Empirical Bayes Thresholding (EBT) in \citet{johnstone2004needles},  and (iv) the Auxiliary Screening (Aux-Scr) procedure using simulated data in Section \ref{simu.sec} and a real dataset in Section \ref{analysis.sec}. The ``Aux-Scr'' method is motivated by a comment for a reviewer. The idea is to first utilize $\bm{S}$ to conduct a preliminary screening of the data, then discard coordinates that appear to contain little information, and finally apply soft-thresholding estimators on remaining coordinates. A detailed description of the Aux-Scr method is provided in Section A of the Supplement.  More simulation results and an additional real data analysis are provided in Sections D and E of the Supplement. Our numerical results
suggest that ASUS enjoys superior numerical performance  and the efficiency gain over competitive  estimators is substantial in many settings.

\subsection{Implementation and \textsf{R}-package \texttt{asus}}

The \textsf{R}-package \href{https://github.com/trambakbanerjee/asus#asus}{\texttt{asus}} has been developed to implement our proposed methodology. In this section, we provide some implementation details upon which our package has been built.

Our scheme for choosing $\mfk{T}$ involves minimizing $S(\mfk{T},\pmb Y,\bm S)$ with respect to $\mfk{T}$. In particular, the optimal $\mfk{T}$ is given by
\begin{equation}
\label{eq:2.5}
\hat{\mfk{T}}=\argmin_{\bm{\tau}\in\Delta_n,t_1,\ldots,t_K\in[0,t_n]}\,S(\mfk{T},\pmb Y,\bm S)
\end{equation}
where $\Delta_n$ is a collection of $K-1$ dimensional distinct points spanning $\mbf{R}^{K-1}_{+}$ and $t_n$ denotes the universal threshold of $\sqrt{2\log n}$. To solve this minimization problem, we proceed as follows: Let ${S}_{(1)},{S}_{(n)}$ be the smallest and largest $S_i$ respectively. Consider a set of $m_n$ equi-spaced points spanning $({S}_{(1)},{S}_{(n)})$ and take $\Delta_n$ to be a $\binom{m_n}{K-1}\times K-1$ matrix where each row is a $K-1$ dimensional sorted vector constructed out of the $m_n$ points. For each $\bm{\tau}^j$ in the $j$th row of $\Delta_n$, determine $\{t_1^j,\ldots,t_K^j\}$ by minimizing the SURE function for the $K$ groups $\widehat{\mc{I}}_k^\tau$. This step can easily be carried out via the hybrid scheme discussed in \cite{donoho1995adapting}. Using Proposition \ref{prop1}, we compute $S(\mfk{T},\pmb Y,\bm S)$ at $\mfk{T}=\{\bm{\tau}^j,t_1^j,\ldots,t_K^j\}$, and repeat this process for $j=1,\ldots,\binom{m_n}{K-1}$ to find $\hat{\mfk{T}}$ using equation \eqref{eq:2.5}. For choosing an appropriate $K$, the procedure discussed above can be repeated for each candidate value of $K$ and an estimate of $K$ may be taken to be the one that minimizes the SURE estimate of risk of ASUS over the candidate values of $K$. In Section \prp{F} of the Supplementary Material, we present a simple example that demonstrates this procedure for choosing $K$. Our practical recommendation is to take $m_n=50\log n$ and $K=2$ which is computationally inexpensive and tends to provide substantial reduction in overall risk against the competing estimators in both simulations and real data examples we considered.

\subsection{Simulation} \label{simu.sec}

This section presents results from two simulation studies, respectively investigating the performances of ASUS in one-sample and two-sample estimation problems. To reveal the usefulness of side information and investigate the effectiveness of ASUS, we also include the oracle estimator $\tilde{\bm{\theta}}^{SI}(\mfk{T}^{OR}_n)$ in the comparison. The MSE of the oracle estimator (OR), which provides the lowest attainable risk, serves as a benchmark for assessing the performance of various methods. The R code that reproduces our simulation results can be downloaded from the following link -- \href{https://github.com/trambakbanerjee/ASUS_paper}{https://github.com/trambakbanerjee/ASUS}.

\subsubsection{One-sample estimation with side information}
\label{sec:simexp1}
\begin{figure}[!h]
	\centering
	\includegraphics[width=1\linewidth]{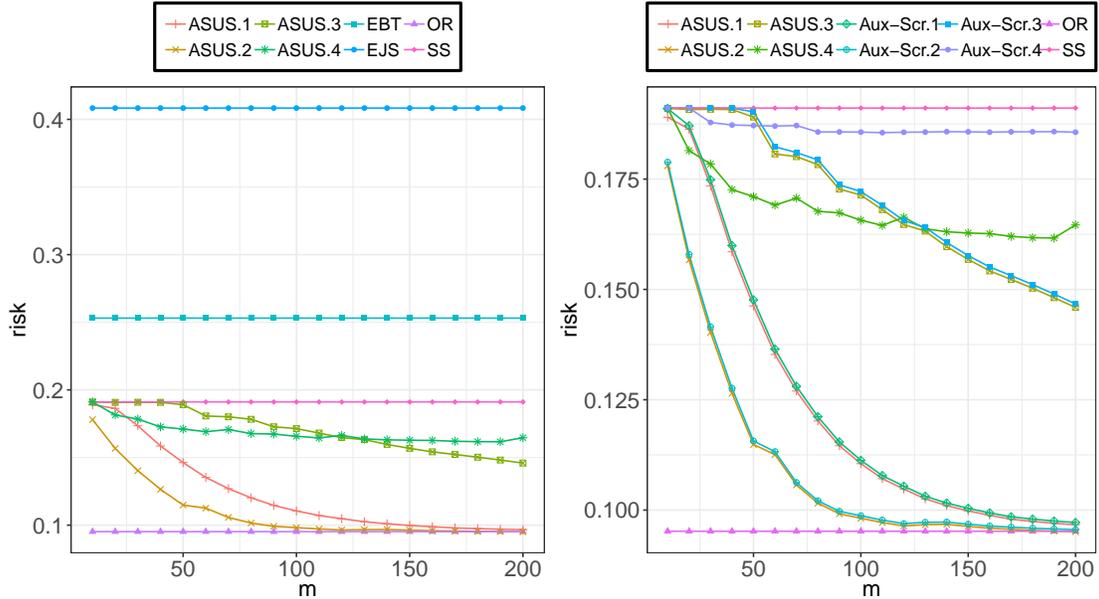}
	\caption{One-sample estimation with side information for scenario S1: Estimated risks of different estimators. Left: ASUS versus EBT and EJS. Right: ASUS verus Aux-Scr.}
	\label{exp1sim1}
\end{figure}

\begin{figure}[!h]
		\centering
		\includegraphics[width=1\linewidth]{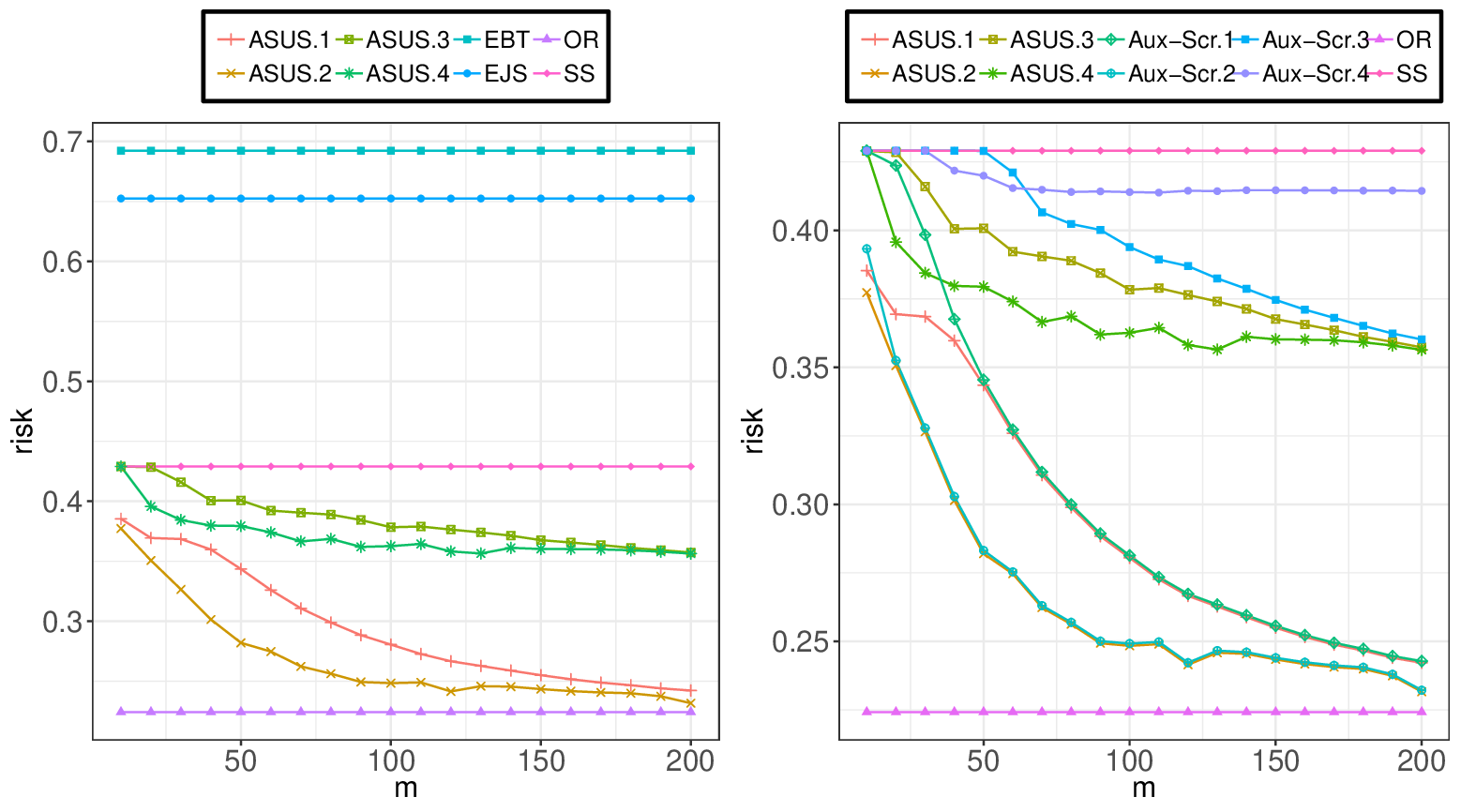}
		\caption{One-sample estimation with side information for scenario S2: Estimated risks of different estimators. Left: ASUS versus EBT and EJS. Right: ASUS verus Aux-Scr.}
\label{exp1sim2}
\end{figure}

We generate our data based on hierarchical Models \eqref{y-theta} to \eqref{s-xi}, where we fix $n=5000$, $K=2$, and take $h_\theta(\xi_i,\eta_{1i}) = {\xi_i}+\eta_{1i}$. We simulate $\eta_{1i}$ from a sparse mixture model $(1-n^{-1/2})\delta_0+n^{-1/2}\mathrm{N}(2,0.01)$. The latent vector $\bm{\xi}$ is simulated under the following two scenarios: 
\begin{itemize}
	\item[(S1)]$\bm{\xi}\sim\Big(\underbrace{\mathrm{Unif}(6,7)}_\text{sample size $=50$},\underbrace{\mathrm{Unif}(2,3)}_\text{sample size $=200$},\underbrace{0,\ldots\ldots,0}_\text{sample size $=n-250$}\Big)$,
	\item[(S2)] $\bm{\xi}\sim\Big(\underbrace{\mathrm{Unif}(4,8)}_\text{sample size $=200$},\underbrace{\mathrm{Unif}(1,3)}_\text{sample size $=800$},\underbrace{0,\ldots\ldots,0}_\text{sample size $=n-10^3$}\Big)$
\end{itemize}
with $Y_i\sim \mathrm{N}(\theta_i,1)$. In practice, we only observe an auxiliary sequence $\bm S$, which can be viewed as a noisy version of $\bm \xi$. To assess the impact of noise on the performance of ASUS, we consider four different settings. In settings 1 and 2, we simulate $m$ samples of $\bm{\eta}_2=(\eta_{21},\dots,\eta_{2n})$ from two different distributions and generate 
\prp{auxiliary sequences} $\bm S_1$ and $\bm S_2$ as follows: 
\begin{description}
 \item (1)\; $\eta_{2i}^{(1)}\stackrel{i.i.d}{\sim}\mathrm{Laplace}(0,4)$ with $\bm S_1=|\bm{\xi}+\bm{\bar{\eta}}_{2}^{(1)}|$,  
 \item (2)\; $\eta_{2i}^{(2)}\stackrel{i.i.d}{\sim}\chi^{2}_{10}$ with $\bm S_2=|\bm{\xi}+\bm{\bar{\eta}}_{2}^{(2)}|$,
\end{description}
where $\bm{\bar{\eta}}_{2}^{(k)}$ is the average of $\bm{\eta}_{2}^{(k)}$ over the $m$ samples. For settings 3 and 4, we first introduce perturbations in the latent variable vector $\bm{\xi}$ and then generate \prp{auxiliary sequences} $\bm S_3$, $\bm S_4$ as follows: 
\begin{description}
\item (3)\; $\tilde{\xi}_{i}=\xi_i\,\mathrm{I}_{\xi_i\ne 0}+\mathrm{LogN}(0,5/\sqrt{m})\,\mathrm{I}_{\xi_i=0}$ with 
\prp{$\bm S_3=|\tilde{\bm{\xi}}+\bm \rho\otimes\bm{\bar{\eta}}_{2}^{(1)}|$, where $\bm \rho$ is a vector of $n$ Rademacher random variables generated independently.}
\item (4)\; $\tilde{\xi}_i=\xi_i\,\mathrm{I}_{\xi_i\ne 0}+\mathrm{t}_{2m/10}\,\mathrm{I}_{\xi_i=0}$ with 
\prp{$\bm S_4=|\tilde{\bm{\xi}}-\bm \rho\otimes\bm{\bar{\eta}}_{2}^{(2)}|$, where $\bm \rho$ is a vector of $n$ independent Bernoulli random variables with probability of success $0.75$.}
\end{description}
We vary $m$ from $10$ to $200$ to investigate the impact of noise. The MSEs are obtained by averaging over $N=500$ replications. The results for scenarios S1 and S2 are summarized in table \ref{tab:simexp1} and in Figures \ref{exp1sim1} and \ref{exp1sim2} wherein ASUS.j and Aux-Scr.j correspond to versions of ASUS and Aux-Scr that rely on the side information in the auxiliary sequence $\bm S_j$, $j=1,\ldots,4$.

\begin{table}[!t]
	\centering
	\caption{One-sample estimation with side information: risk estimates and estimates of $\mfk{T}$ for ASUS at $m=200$. Here $n_k^{^\star}=|\mc{I}_{k}^{\tau^\star}|$ and $n_k=|\widehat{\mc{I}}_{k}^{\tau}|$ for $k=1,2$.}
	\scalebox{0.8}{
	\begin{tabular}{clcc}
		\toprule
		&       & \multicolumn{2}{c}{One-sample estimation with side information} \\
		\midrule
		&       & Scenario S1    & Scenario S2 \\
		\midrule
		\multirow{4}[0]{*}{OR} & $\tau^\star$     & 2     & 1.003 \\
		& $t_1^\star$, $t_2^\star$ & 4.114, 0.138 & 4.073, 0.133 \\
		& $n_1^\star$, $n_2^\star$ &  4750, 250  & 4008, 992 \\
		& risk  & 0.095 & 0.224 \\
		\midrule
		\multirow{4}[0]{*}{ASUS.1} & $\tau$   & 1.342 & 0.979 \\
		& $t_1$, $t_2$ & 4.114, 0.107 & 4.073, 0.156 \\
		& $n_1$, $n_2$ &   4748, 252    & 4008, 992	 \\
		& risk  & 0.097 & 0.243 \\
		\midrule
		\multirow{4}[0]{*}{ASUS.2} & $\tau$   & 11.229 & 5.82 \\
		& $t_1$, $t_2$ & 4.115, 0.106 & 4.073, 0.137 \\
		& $n_1$, $n_2$ &  4748, 252  & 4008, 992 \\
		& risk  & 0.095 & 0.228 \\
		\midrule
		\multirow{4}[0]{*}{ASUS.3} & $\tau$   & 1.777 & 1.778 \\
		& $t_1$, $t_2$ & 4.089, 0.662 & 3.422, 0.441 \\
		& $n_1$, $n_2$ &  4271, 729 & 3606, 1394 \\
		& risk  & 0.146 & 0.357 \\
		\midrule
		\multirow{4}[0]{*}{ASUS.4} & $\tau$   & 7.785 & 8.524 \\
		& $t_1$, $t_2$ & 1.360, 3.653 & 0.745, 3.864 \\
		& $n_1$, $n_2$ &  1775, 3225 & 2249, 2751 \\
		& risk  & 0.165 & 0.356 \\
		\midrule
		Aux-Scr.1 & risk  & 0.097 & 0.243 \\
		Aux-Scr.2 & risk  & 0.095 & 0.232 \\
		Aux-Scr.3 & risk  & 0.147 & 0.360 \\
		Aux-Scr.4 & risk  & 0.186 & 0.414 \\
		\midrule
		SureShrink & risk  & 0.191 & 0.429 \\
		EBT   & risk  & 0.253 & 0.692 \\
		EJS   & risk  & 0.408 & 0.652 \\
		\bottomrule
	\end{tabular}}
	\label{tab:simexp1}%
\end{table}%

\prp{From the left panels of figures \ref{exp1sim1} and \ref{exp1sim2} we} see that ASUS exhibits the best performance when compared against EBT, EJS and SureShrink estimators. 
In particular, ASUS.1, ASUS.2 outperform their counterparts ASUS.3, ASUS.4. This reveals how the usefulness of the latent sequence $\bm\xi$ would affect the performance of ASUS. Nonetheless, ASUS.3 and ASUS.4 still provide improvements over, and, crucially, are never worse than the SureShrink estimator. 
This reveals the impact of the accuracy of the auxiliary sequence $\bm S$ (in capturing the information in $\bm\xi$) on the performance of ASUS. \prp{The right panels of figures \ref{exp1sim1} and \ref{exp1sim2} present the risk comparison between ASUS and Aux-Scr using the auxiliary sequences $\bm S_1,\ldots,\bm S_4$. Not surprisingly, ASUS and Aux-Scr have almost identical risk performance using the auxiliary sequences $\bm S_1,\bm S_2$ and $\bm S_3$ for large $m$. As $m$ increases, the accuracy of these auxiliary sequences increase but the negative Bernoulli perturbations in $\bm S_4$ interferes with its magnitude so that a smaller $|S_{i4}|$ may correspond to a signal coordinate. The Aux-Scr procedure which discards observations based on the magnitude of the auxiliary sequence may miss important signal coordinates while relying on $\bm S_4$. ASUS, however, does not discard any observations and continues to exploit the available information in the noisy auxiliary sequences.}

In table \ref{tab:simexp1}, we report risk estimates and estimates of $\mfk{T}$ for ASUS when $m=200$. \prp{The estimates of the hyper-parameters of Aux-Scr are provided in table 2 of the supplementary material and we only report its risk estimates here in table \ref{tab:simexp1}.} We can see that ASUS.1 and ASUS.2 choose similar thresholding hyper-parameters ($t_1, t_2$) as those of the oracle estimator. \prp{Moreover, ASUS.4 demonstrates a lower estimation risk than Aux-Scr.4 using the same auxiliary sequence $\bm S_4$.}

\subsubsection{Two-sample estimation with side information}
\begin{figure}[!t]
	\centering
	\includegraphics[width=1\linewidth]{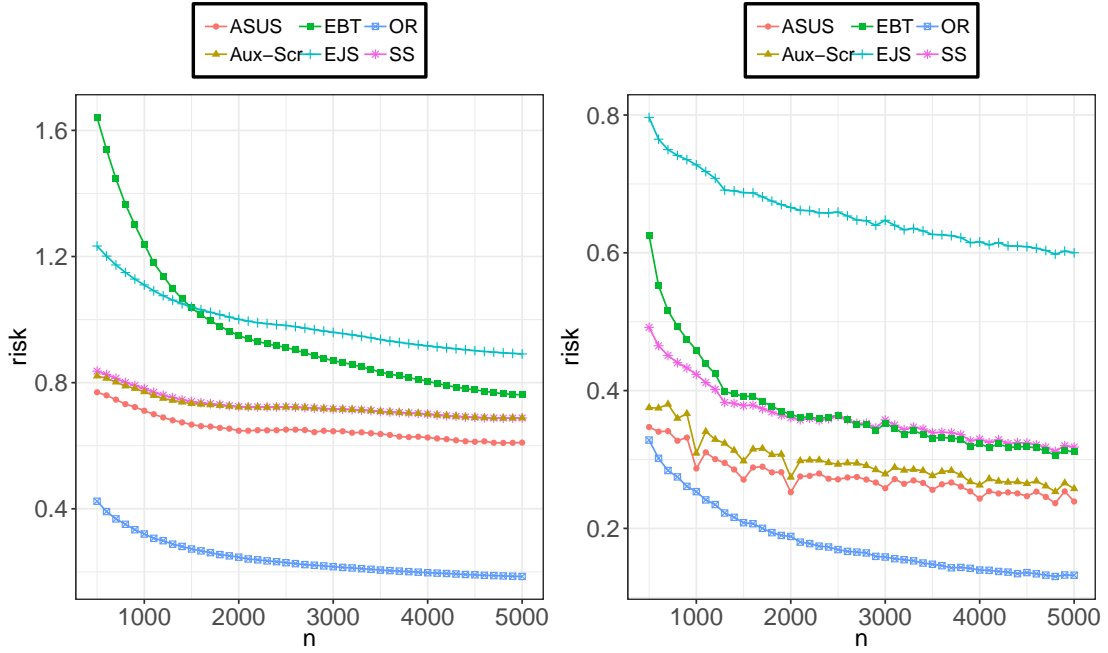}
	\caption{Two-sample estimation with side information: Average risks of different estimators. Left: Scenario S1 and Right: Scenario S2.}
	\label{simexp2}
\end{figure}

We consider the problem of estimating the difference of two Gaussian mean vectors. An auxiliary sequence can be constructed from data by following Example 3 in Section \ref{principles.subsec}. We first simulate 
$$\xi_{1i}\sim~(1-p_1)\delta_0+p_1\,\mathrm{Unif}(3,7), \quad \xi_{2i}\sim~(1-p_2)\delta_0+p_2\,\delta_{\{4\}},
$$
where $\delta_{\{4\}}$ is the dirac delta at $4$ and then generate $\mu_{i,1}=\xi_{1i}+\eta_{1i}$ and $\mu_{i,2}=\xi_{2i}+\eta_{2i}$ with $\eta_{1i},\eta_{2i}\stackrel{i.i.d}{\sim}N(0,0.01)$. The parameter of interest is $\bm \theta = \bm \mu_1-\bm \mu_2$ and the associated latent side information vector is $\bm \xi=\bm \xi_1-\bm \xi_2$. The observations based on the simulated mean vectors are generated as $U_i\sim~\mathrm{N}(\mu_{i,1},\sigma_{i,1}^2), \quad V_i\sim~ \mathrm{N}(\mu_{i,2},\sigma_{i,2}^2)$. Finally, the primary and auxiliary statistics are obtained as $Y_i=U_i-V_i, \quad S_i=|U_i+\kappa_i V_i|$. We fix $p_1 = n^{-0.6}$, $p_2=n^{-0.3}$, $\kappa_i=\sigma_{i,1}/\sigma_{i,2}$ and consider two scenarios where $\sigma_{i,1}=\sigma_{i,2}=1$ under scenario S1 and $(\sigma_{1,i}^2,\sigma_{2,i}^2)\stackrel{i.i.d}{\sim} \mathrm{Unif}(0.1,1)$ under scenario S2. The estimates of risks are obtained by averaging over $N=1000$ replications. We vary $n$ from $500$ to $5000$ to investigate the impact of the strength of side information. 
The simulation results are reported in Table \ref{tab:simexp2} and figure \ref{simexp2}.

We see that ASUS uses the side information in $\bm S$ and exhibits the best performance across both scenarios. In scenario S2, the variances of $Y_i$ are smaller, which leads to an improved risk performance of ASUS over scenario S1. Similar to the previous simulation study, the risk of ASUS would not exceed the risk of the SureShrink estimator across both the scenarios. Different magnitudes of the thresholding hyper-parameters $(t_1,t_2)$ in table \ref{tab:simexp2} further corroborates the importance of the auxiliary statistics $S_i$ in constructing groups with disparate sparsity levels and thereby improving the overall estimation accuracy. This is particularly true in the case of scenario S2 where EBT and SureShrink are competitive but ASUS is far more efficient because it has constructed two groups where one group holds majority of the signals and ASUS uses the smaller threshold $t_2$ to retain the signals. The other group holds majority of the noise wherein ASUS uses the larger threshold $t_1$ to shrink them to zero. \prp{Moreover, we notice that ASUS provides a better risk performance than Aux-Scr across both the scenarios. Using the side information in $\bm S$, Aux-Scr discards observations that have $|S_{i}|\le \tau$ thereby eliminating some potentially information rich signal coordinates and thus returns a higher risk than ASUS.}
\begin{table}[!t]
	\centering
	\caption{Two-sample estimation with side information: risk estimates and estimates of $\mfk{T}$ for ASUS at $n=5000$. Here $n_k^{^\star}=|\mc{I}_{k}^{\tau^\star}|$ and $n_k=|\widehat{\mc{I}}_{k}^{\tau}|$ for $k=1,2$.}
	\scalebox{0.8}{
		\begin{tabular}{clcc}
			\toprule
			&       & \multicolumn{2}{c}{Two-sample estimation with side information} \\
			\midrule
			&       & Scenario S1    & Scenario S2 \\
			\midrule
			\multirow{4}[0]{*}{OR} & $\tau^\star$   & 1.947 & 1.363 \\
			& $t_1^\star$, $t_2^\star$ & 4.106, 0.137 & 4.106, 0.424 \\
			& $n_1^\star$, $n_2^\star$ & 4584,  416 & 4583, 417\\
			& risk  & 0.185 & 0.132 \\
			\midrule
			\multirow{4}[0]{*}{ASUS} & $\tau$   & 3.167 & 2.504 \\
			& $t_1$, $t_2$ & 1.223, 0.253 & 3.058, 0.323 \\
			& $n_1$, $n_2$ & 4570, 430 & 4195, 805\\
			& risk  & 0.610 & 0.239 \\
			\midrule
			\multirow{4}[0]{*}{Aux-Scr} & $\tau$   & 14.385 & 2.768 \\
			& $t_1$, $t_2$ & 0.955, 0.002 & 5.708, 0.498 \\
			& $n_1$, $n_2$ & 4991, 9 & 3681, 1319\\
			& risk  & 0.688 & 0.258 \\
			\midrule
			SureShrink & risk  & 0.688 & 0.318 \\
			EBT   & risk  & 0.761 & 0.311 \\
			EJS   & risk  & 0.891 & 0.600 \\
			\bottomrule
	\end{tabular}}
	\label{tab:simexp2}%
\end{table}%

\subsection{Analysis of RNA sequence data}
\label{analysis.sec}

We compare the performance of ASUS against the SureShrink (SS) estimator for analysis of the RNA sequence data described in the introduction. The goal is to estimate the true expression levels $\bm{\theta}$ of the $n$ genes that are infected with VZV strain. {Through previous studies conducted in the lab, expression levels corresponding to other four experimental conditions, including uninfected cells  (C1, 3 replicates), a fibrosarcoma cell line (C2, 3 replicates) and cells treated with interferons gamma (C3, 2 replicates), alpha (C4, 3 replicates), were also collected.} Let $X_i$ be the mean expression level of gene $i$ across the four experimental conditions. Set $S_i = |X_i|$ with $K=2$. Let $\hat{\theta}_i^S(t)$ denote the SureShrink estimator of $\theta_i$ based on $Y_i$, the mean expression level of gene $i$ under the VZV condition. The standard deviation $\sigma_i$ for the mean expression level pertaining to gene $i$ across the 3 replicates of the VZV strain is derived from the study conducted in \cite{sen2018distinctive}.

\begin{figure}[!h]
	\centering
	\includegraphics[width=1\linewidth]{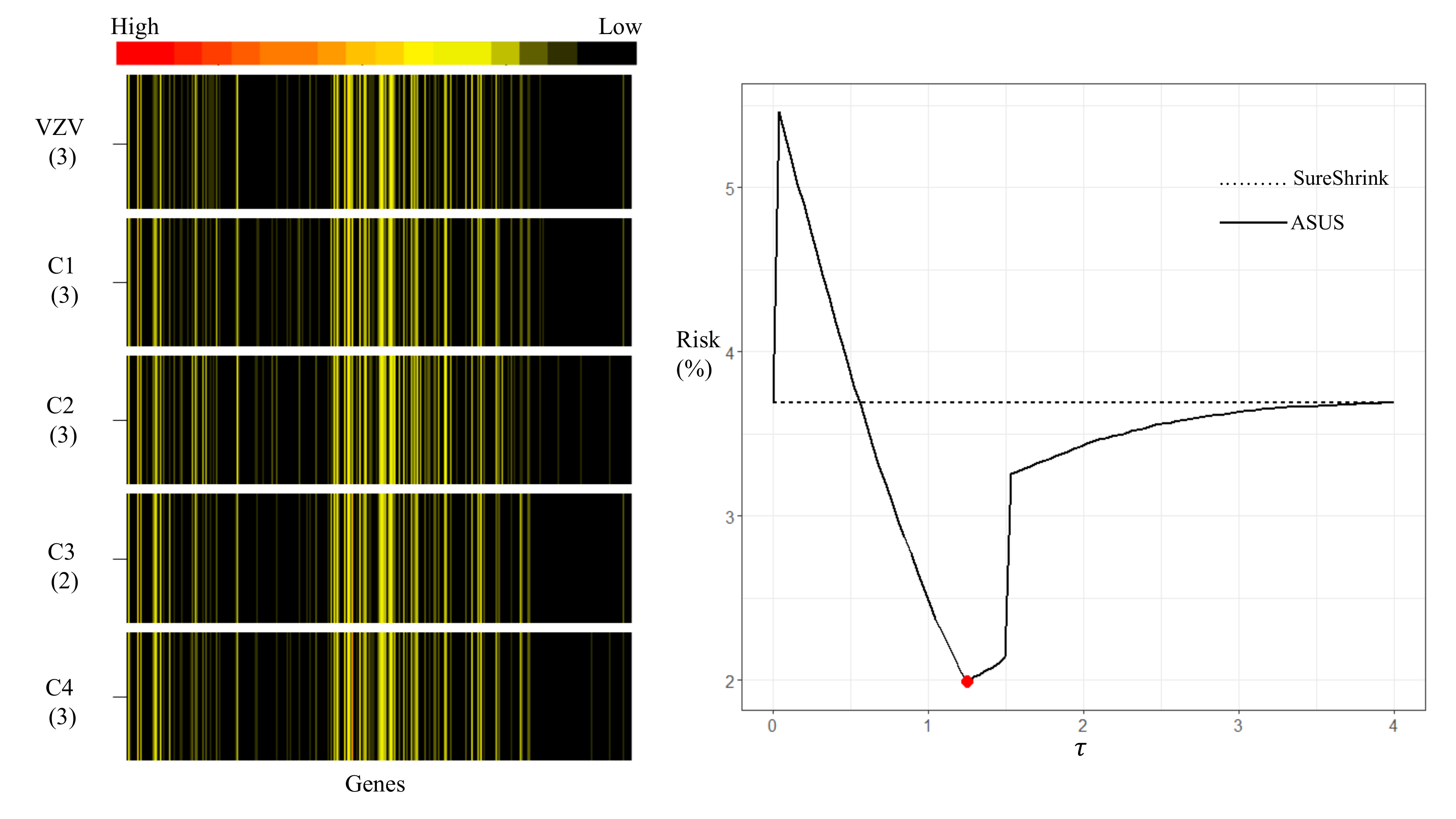}
	\caption{Left: Heat map showing the following from top to bottom: average expression levels of VZV, C1, C2, C3 and C4 across  their respective replicates (in parenthesis). Right: SURE estimate of the risk of $\hat{\theta}^{S}_i(t)$ at $t=0.61$ versus an unbiased estimate of the risk of ASUS for different values of $\tau$.}
	\label{fig:1}
\end{figure}

{On the right panel of Figure \ref{fig:1}, the dotted line represents the minimum of the SURE risk of $\hat{\bm{\theta}}^S(t)$, which is minimized at $t=0.61$. The solid line represents the minimum of the SURE risk of a class of two-group estimators over a grid of $\tau$ values. ASUS chooses $\tau$ that minimizes the SURE risk (the red dot in figure \ref{fig:1}). The resulting risk is $1.99\%$ at $\hat{\mfk{T}}=(1.25,1.16,0)$, a significant reduction compared to the risk estimate of $3.69\%$ for $\hat{\bm{\theta}}^S(t)$. In order to evaluate the results in a predictive framework, we next use only two replicates of the VZV strain for calibrating the hyper-parameters and calculate the prediction errors based on the hold out third replicate. The risk reduction by ASUS over SureShrink is about $30\%$.}

\begin{figure}[!t]
	\centering
	\begin{subfigure}{.45\textwidth}
		\centering
		\includegraphics[width=1\linewidth]{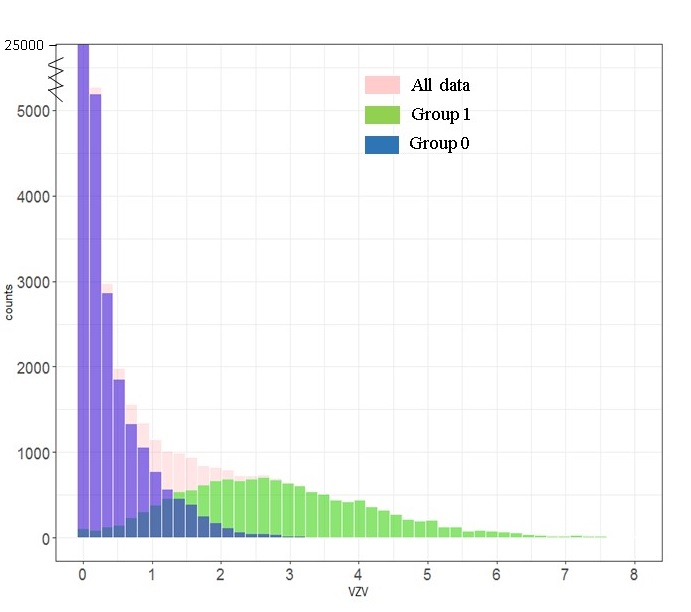}
		\caption{}
		\label{rna1}
	\end{subfigure}%
	\begin{subfigure}{.5\textwidth}
		\centering
		\includegraphics[width=1\linewidth]{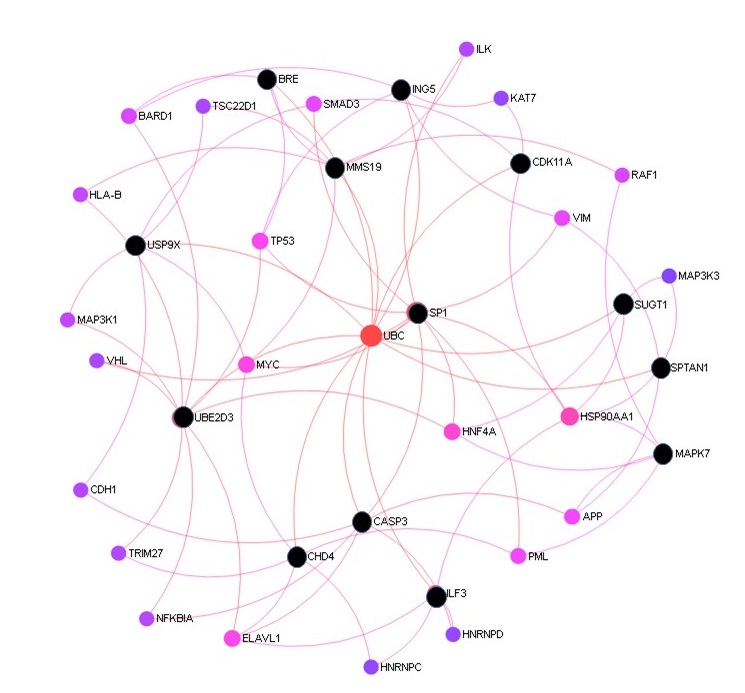}
		\caption{}
		\label{rna2}
	\end{subfigure}
	\caption{(a) Histogram of gene expressions for VZV. Group 1 is $\widehat{\mc{I}}^{\tau}_2$ and Group 0 is $\widehat{\mc{I}}^{\tau}_1$. (b) A network of 20 new genes highlighted in black with their interaction partners.}
	\label{}
\end{figure}

In this example, a reduction in risk is possible because ASUS has efficiently exploited the sparsity information about $\bm{\theta}$ encoded by $\bm S$. This can be seen, for example, from (i) the stark contrast between the magnitudes of thresholding hyper-parameters $t_1, t_2$ for the two groups in table \ref{tab:rna} and (ii) the heat maps in figure \ref{fig:1} where the genes expressions under the four experimental conditions follow the expression pattern of VZV. \prp{Moreover, the risk of Aux-Scr for this example was seen to be no better than the SureShrink estimator and thus has been excluded from the results reported in table \ref{tab:rna}.} Figure \ref{rna1} presents the distribution of gene expression for genes that belong to  groups $\widehat{\mc{I}}_1^\tau$ and $\widehat{\mc{I}}_2^\tau$. ASUS exploits the side information in $\bm S$ to partition the estimation units into two groups with very different sparsity levels and therefore returns a much smaller risk.

\begin{table}[!h]
	\centering
	\caption{Summary of SureShrink and ASUS methods (RNA-Seq data). $n_k=|\widehat{\mc{I}}_{k}^{\tau}|$ for $k=1,2$.}
	\scalebox{0.8}{\begin{tabular}{ccc}
		\toprule
		&       & RNA Seq \\
		\midrule
		& $n$     & 53,216 \\ 
		\midrule
		\multirow{2}[0]{*}{SureShrink} & $t$     & 0.61 \\
		& SURE estimate & 3.69\\
		\midrule
		\multirow{6}[0]{*}{ASUS} & $\tau$   & 1.25 \\
		& $t_1$  & 1.16\\
		& $t_2$  & 0 \\
		& $n_1$  & 39,535\\
		& $n_2$  & 13,681\\
		& SURE estimate & 1.99 \\
		\bottomrule
	\end{tabular}}%
	\label{tab:rna}%
\end{table}%
The ASUS estimator $\hat{\bm{\theta}}^{SI}(\hat{\mfk{T}})$ results in the discovery of $114$ new genes than those discovered by using $\hat{\bm{\theta}}^{SI}(t)$. Figure \ref{rna2} shows the network of protein-protein interactions of 20 such genes. The interaction network is generated using NetworkAnalyst \citep{xia2015networkanalyst} that maps the chosen genes to a comprehensive high-quality protein-protein interaction (PPI) database based on InnateDB. A search algorithm is then performed to identify first-order neighbors (genes that directly interact with a given gene) for each of these mapped genes. The resulting nodes and their interaction partners are returned to build the network. In case of the RNA-Seq data, the interaction network of the 20 new genes indicates that ASUS may help reveal important biological synergies between genes that have a high estimated expression level for VZV and other genes in the human genome.

\section{Discussion}
\label{sec:discuss}

In high-dimensional estimation and testing problems, the sparsity structure can be encoded in various ways; we have considered three basic settings where the structural information on sparsity may be extracted from (i) prior or domain-specific knowledge, (ii) covariate sequence based on the same data, or (iii) summary statistics based on secondary data sources. This article develops a general integrative framework for sparse estimation that is capable of handling all three scenarios. We use higher-order minimax optimality tools to establish the adaptivity and robustness of ASUS. Numerical studies using both simulated and real data corroborate the improvement of ASUS over existing methods. 

We conclude the article with a discussion of several open issues. Firstly, in large-scale compound estimation problems, various data structures such as sparsity, heteroscedasticity, dependency and hierarchy are often available alongside the primary summary statistics. ASUS can only handle the sparsity structure; and it is desirable to develop a unified framework that can effectively incorporate other types of structures into inference. New theoretical frameworks will be needed to characterize the usefulness of various types of side information and to establish precise conditions under which the new integrative method is asymptotically optimal. Secondly, in situations where there are multiple auxiliary sequences, it is unclear how to modify the ASUS framework to construct groups using an auxiliary matrix. The computation involved in the search for the optimal group-wise thresholds, which requires the evaluation of the SURE function for every possible combination of group-wise thresholds, quickly becomes prohibitively expensive as the number of columns increases. Finally, the higher dimension would affect the stability of an integrative procedure adversely. A promising idea for handling multiple auxiliary sequences is to construct a new auxiliary sequence that represents the ``optimal use'' of all available side information. However, the search for this optimal direction of projection is quite challenging. It would be of great interest to explore these directions in future research. 

\subsection*{Acknowledgments}

We thank Ann Arvin and Nandini Sen for helpful discussions on the virology application. We thank the AE and two referees for the constructive suggestions that have greatly helped to improve the presentation of the paper. In particular, we are grateful to an excellent comment from a referee that leads to the Bayesian interpretation of ASUS in Section \ref{sec:empbayes}.

\bibliographystyle{chicago}
\bibliography{paperref}

\end{document}


\title{Supplementary Material for ``Adaptive Sparse Estimation with Side Information''}
	\date{}
	\maketitle
This supplement contains a detailed description of the Auxiliary Screening procedure (Aux-Scr) (Section \ref{sec:aux-scr}), proofs of the results in Section 2 and 3 of the main paper (Sections \ref{supp:moreproofs_2} and \ref{supp:moreproofs_3} respectively), additional simulation experiments (Section \ref{supp:moresims}), a real data analysis (Section \ref{supp:mtc}) and an example that demonstrates a data driven procedure for choosing $K$ (Section \ref{supp:chooseK}).
\appendix
\section{The Auxiliary Screening approach}
\label{sec:aux-scr}
We consider a potential competitor of ASUS, called Aux-Scr, which uses the auxiliary sequence $\bm{S}$ to conduct a preliminary screening of the primary data thereby discarding data instances that contain little information and retains the potentially information rich primary data for estimation. Using the notation described in the main paper, we define Aux-Scr for $K=2$ groups as follows:

Let $\bm Y$ and $\bm S$ denote the primary statistics and auxiliary sequence obeying the models (1)-(3) described in Section 2.1 of the main paper. Let $\eta_{t}(.)$ be a soft-thresholding operator such that
\begin{equation*}
\eta_{t}(Y_i)=\begin{cases}
-Y_i\sigma^{-1}_i\text{, if }|Y_i\sigma^{-1}_i|\le t;\\
-t\text{ sign}(Y_i\sigma^{-1}_i)\text{, otherwise}.
\end{cases}
\end{equation*}
The Aux-Scr estimator operates in two steps: first it constructs $K=2$ groups using the magnitude of $\bm S$, where in group 1 $|S_i|\le\tau$ and in group 2 $|S_i|> \tau$. Then it conducts soft-thresholding estimation using the primary statistics $\bm Y$ in group 2 and estimates $\widehat{\theta}_i=0$ for all coordinates that belong to group 1. The tuning parameters for both grouping and shrinkage are determined using the SURE criterion. 
\begin{proc} 
	For $k=1,2$, denote $\widehat{\mc{I}}_1^\tau=\{i:|{S}_i|\le\tau\}$ and $\widehat{\mc{I}}_2^\tau=\{i:|{S}_i|>\tau\}$. Consider the following class of shrinkage estimators:
	\begin{equation*}
	\hat{\theta}^{SI}_{i}(\mfk{T})\coloneqq Y_i+\sigma_i \eta_{t_k}(Y_i)\text{ if }i\in\widehat{\mc{I}}_k^\tau,
	\end{equation*}
	where, $\mfk{T}=\{\tau,t_2\}$,   
	$t_2$ varies in $[0,t_n]$ with $t_n=(2\log n)^{1/2}$ and $t_1=\max\{|Y_i\sigma_i^{-1}|:i\in \widehat{\mc{I}}_1^\tau\}$. Thus, the set of all possible hyper-parameter $\mfk{T}$ values is $\mathcal{H}_n= \mbf{R}_+\times[0,t_n]$. Define the SURE function
	\begin{equation}
	\label{eq:auxscr-sure}
	S(\mfk{T},\bm Y,\bm S)=n^{-1}\left[\sum_{i=1}^{n}\sigma^2_i+\sum_{k=1}^{K}	\sum_{i\in \widehat{\mc{I}}_k^\tau} \left\{\sigma^2_i(|Y_i\sigma_i^{-1}|\wedge t_k)^{2}
	-2\sigma^2_i\,I(|Y_i\sigma_i^{-1}|\le t_k)\right\}\right].
	\end{equation}
	Let $\hat{\mfk{T}}=\argmin_{\mfk T \in \mathcal{H}_n}\,S(\mfk{T},\bm Y,\bm S)$. Then, the Aux-Scr estimator is given by $\hat{\theta}^{SI}_{i}(\hat{\mfk{T}})$ with $t_1=\max\{|Y_i\sigma_i^{-1}|:i\in \widehat{\mc{I}}_1^\tau\}$.
\end{proc}
Following the arguments in the proof of Proposition 1, it can be shown that equation \eqref{eq:auxscr-sure} is an unbiased estimate of the true risk. Moreover, unlike ASUS, the thresholding hyper-parameter $t_1$ is fixed and the SURE criteria is used to select the grouping hyper-parameter $\tau$ and the thresholding hyper-parameter $t_2$. 

When compared to Aux-Scr, ASUS has three distinct advantages: optimality, robustness and adaptivity. First, the screening strategy does not address the important issue on how to set an optimal group-wise cutoff in the screening stage; this issue has been resolved by the SURE criterion in ASUS. Second, the ``divide-and-threshold" strategy adopted by ASUS is clearly more effective than the ``screening" strategy that directly throws away a lot of data. When $\bm S$ is imperfect in capturing the sparsity structure, the screening step would inevitably miss important signal coordinates. By contrast, ASUS is more robust to noisy side information as it only utilizes $\bm S$ to divide $\bm Y$ into groups; no coordinates are discarded directly. Finally, Aux-Scr uses the same threshold for all coordinates that pass the preliminary screening stage. By contrast, ASUS is more adaptive to the unknown sparsity as it sets varied group-wise thresholds to reflect the possibly varied sparsity levels across groups.

\section{Proofs of the results in Section 2}
\label{supp:moreproofs_2}
\textbf{Proof of Proposition 1} - Recall that $r_n(\mfk{T};\bm{\theta})=n^{-1}\mbb{E}||\hat{\bm{\theta}}^{SI}(\mfk{T})-\bm{\theta}||^2$ where the expectation is taken with respect to the joint distribution of $(Y_i,S_i)$ given $\xi_i$ for $i=1,2,\cdots,n$. Now, expanding $||\hat{\bm{\theta}}^{SI}(\mfk{T})-\bm{\theta}||^2$ as $||\bm{Y}-\bm{\theta}||^2+||\hat{\bm{\theta}}^{SI}(\mfk{T})-\bm{Y}||+2\langle \bm{Y}-\bm{\theta},\hat{\bm{\theta}}^{SI}(\mfk{T})-\bm{Y}\rangle$ and taking expectation, we have,
\begin{equation*}
n\,r_n(\mfk{T};\bm{\theta})=\sum_{i=1}^{n}\sigma_i^2+\sum_{i=1}^{n}\sum_{k=1}^{K}\sigma^2_i\mbb{E}\Bigl\{\eta^2_{t_k}(Y_i)I(i\in\mc{I}_k^\tau)\Bigr\}+2\sum_{i=1}^{n}\sum_{k=1}^{K}\sigma_i\mbb{E}\Bigl\{(Y_i-\theta_i)\eta_{t_k}(Y_i)I(i\in\mc{I}_k^\tau)\Bigr\}
\end{equation*}
Observe that from Models (1) to (3), the pairs $\Bigl\{(Y_i-\theta_i)\eta_{t_k}(Y_i),~ I(i\in\widehat{\mc{I}}_k^\tau)\Bigr\}$  are uncorrelated for each $i$ and for all $k=1,\ldots,K$. Further, note that by Lemma~1 of \citet{stein1981estimation}
\begin{eqnarray*}
	\mbb{E}\Bigl\{\eta^{'}_{t_k}(Y_i)\Bigr\}&=&\sigma_i^{-1}\int_{\mbf{R}}^{}\eta^{'}_{t_k}(u)\phi\Big(\dfrac{u-\theta_i}{\sigma_i}\Big)du = \sigma_i^{-2}\mbb{E}\Bigl\{\eta_{t_k}(Y_i)(Y_i-\theta_i)\Bigr\}~.
\end{eqnarray*}
Thus, $-\sigma^2_{i}\mbb{E}\bigl\{I(|Y_i\sigma_i^{-1}|\le t_k)\bigr\}=\sigma_i\mbb{E}\bigl\{\eta_{t_k}(Y_i)(Y_i-\theta_i)\bigr\}$ which completes the proof.
\medskip\\
\noindent\textbf{Proof of Theorem 1, statement (a)} - First note that we can decompose $S(\mfk{T},\pmb Y,\bm S)$ into $K$ components: $S(\mfk{T},\pmb Y,\bm S)=\sum_{k=1}^{K}S_k(\mfk{T}_k,\pmb Y,\bm S)$ where
\begin{equation*}
S_k(\mfk{T}_k,\pmb Y,\bm S)=n^{-1}\sum_{i=1}^{n}\Bigl\{\sigma_i^2-2\sigma_i^2I(|Y_i\sigma_i^{-1}|\le t_k)+\sigma_i^2\big(|Y_i\sigma_i^{-1}|\wedge t_k\big)^2\Bigr\}I\Bigl\{S_i\in(\tau_{k-1},\tau_k]\Bigr\}
\end{equation*}
and $\mfk{T}_k=\{\tau_{k-1},\tau_k,t_k\}$. Let
\begin{equation*}
S_{i}^k(\mfk{T}_k,Y_i,S_i)=\sigma_i^2-2\sigma_i^2I(|Y_i\sigma_i^{-1}|\le t_k)+\sigma_i^2\big(|Y_i\sigma_i^{-1}|\wedge t_k\big)^2I\Bigl\{S_i\in(\tau_{k-1},\tau_k]\Bigr\}
\end{equation*}
and notice that $S_{i}^k(\mfk{T}_k,Y_i,S_i)$ is bounded above by $\sigma_i^2(1+t_k^2)$ for each $i$. Also, we can decompose the risk  $r_n(\mfk{T},\bm{\theta})=\sum_{k=1}^{K}r_n^k(\mfk{T}_k,\bm{\theta})$ where
\begin{equation*}
r_n^k(\mfk{T}_k,\bm{\theta})=n^{-1}\sum_{i=1}^{n}\mbb{E}\Big[\Big(Y_i+\sigma_i\eta_{t_k}(Y_i)-\theta_i\Big)I\Bigl\{X_i\in(\tau_{k-1},\tau_k]\Bigr\}\Big]^2=n^{-1}\sum_{i=1}^{n}r_{i}^k(\mfk{T}_k,\theta_i)
\end{equation*}
and noting that $r_{i}^k(\mfk{T}_k,\theta_i)\le 2\sigma_i^2(1+t_k^2)$. 
The last inequality follows from the upper bound on the risk of soft thresholding estimator at threshold $t_k$. Now, by triangle inequality, it is enough to show
\begin{equation}
\label{eq:1a.1}
c_n\mbb{E}\Bigl\{\sup_{\mfk{T}_k\in\mbf{R}^2\times[0,t_n]}\Big|S_k(\mfk{T}_k,\pmb Y,\bm S)-r_n^k(\mfk{T}_k;\bm{\theta})\Big|\Bigr\}<\infty\text{ for all }i\text{ and for all large }n
\end{equation}
Based on the form of $S_k(\mfk{T}_k,\pmb Y,\bm S)$, we consider a re-parametrization of the problem with respect to $0\le\tilde{\tau}_{k-1}<\tilde{\tau}_k\le 1$ where $\tilde{\tau}_k=\max_{i\in \mc{I}_k^{\tau}}F_{S_i}(\tau_k)$, $\tilde{\tau}_{k-1}=\min_{i\in \mc{I}_k^{\tau}}F_{S_i}(\tau_{k-1})$ and $F_{S_i}$ is the distribution function of $S_i$. The only $\tau_{k-1},\tau_k$ dependent quantity in the expression of
$S_k(\mfk{T}_k,\pmb Y,\bm S)$ is $\widehat{\mc{I}}_k^\tau=\{i:\tau_{k-1}<S_i\le\tau_k\}$ which is re-parametrized to $\widehat{\mc{I}}_k^{\tilde{\tau}}=\{i:\tilde{\tau}_{k-1}<F_{S_i}(s_i)\le\tilde{\tau}_k\}$. This facilitates the analysis since now the supremum with respect to $\tilde{\mfk{T}}_k=\{\tilde{\tau}_{k-1},\tilde{\tau}_k,t_k\}$ is actually over a compact set. 

We will mimic the proof of Proposition 1 of \citet{donoho1995adapting}, hereafter referred to as DJ95P1, to prove equation \eqref{eq:1a.1}. For the other terms similar arguments will continue to hold. 

Let $S_k(\tilde{\mfk{T}}_k,\pmb Y,\bm S)-r_n(\tilde{\mfk{T}}_k,\bm{\theta})=n^{-1}\sum_{i=1}^{n}U_i(\tilde{\mfk{T}}_k)=V_n(\tilde{\mfk{T}}_k)$ where $\mbb{E}\,U_i(\tilde{\mfk{T}}_k)=0$ and from the upper bounds on $S_{i}^k(\mfk{T}_k,Y_i,S_i)$ and $r_{i}^k(\mfk{T}_k,\theta_i)$, $|U_i(\tilde{\mfk{T}}_k)|\le 3(1+t_n^2)\sigma_i^2$. Now replace $Z_n(t)$ in DJ95P1 with $V_n(\tilde{\mfk{T}}_k)$ and notice that Hoeffding's inequality gives, for a fixed $\tilde{\mfk{T}_k}$ and (for now) arbitrary $r_n>1$,
\begin{equation*}
\mbb{P}\Bigl\{\Big|V_n(\tilde{\mfk{T}}_k)\Big|>\dfrac{r_n}{\sqrt{n}}\Bigr\}\le 2\exp{\Bigl\{-\dfrac{nr_n}{18(1+t_n^2)^2\sum_{i=1}^{n}\sigma_i^4}\Bigr\}}
\end{equation*}
Next, for a perturbation $\tilde{\mfk{T}}_k^\prime=\{\tilde{\tau}_{k-1}^\prime,\tilde{\tau}_k^\prime,t_k^\prime\}$ of $\tilde{\mfk{T}}_k$ where $\tilde{\tau}_k^\prime>\tilde{\tau}_k$, $\tilde{\tau}_{k-1}^\prime>\tilde{\tau}_{k-1}$ and $t_k^\prime>t_k$, we wish to bound the increments $\Big|V_n(\tilde{\mfk{T}}_k)-V_n(\tilde{\mfk{T}}_k^\prime)\Big|$. To that effect, define $\tilde{\mfk{T}}_k^{(r,s)}$ to be $\tilde{\mfk{T}}_k$ but with components $(r,s)$ replaced by the components $(r,s)$ of $\tilde{\mfk{T}}_k^\prime$, $r<s=1,2,3$. {Then we can write, dropping the subscript $k$ from $\tilde{\mfk{T}}_k$ for brevity, that
$|S_k(\tilde{\mfk{T}},\pmb Y,\bm S)-S_k(\tilde{\mfk{T}}^\prime,\pmb Y,\bm S)|$ is bounded above by the sum of three terms: 
 $|S_k(\tilde{\mfk{T}},\pmb Y,\bm S)-S_k(\tilde{\mfk{T}}^{(3)},\pmb Y,\bm S)|$,
$|S_k(\tilde{\mfk{T}}^{(3)},\pmb Y,\bm S)-S_k(\tilde{\mfk{T}}^{(2,3)},\pmb Y,\bm S)|$ and
$|S_k(\tilde{\mfk{T}}^{(2,3)},\pmb Y,\bm S)-S_k(\tilde{\mfk{T}}^\prime,\pmb Y,\bm S)|$.
The first term is bounded by $n^{-1}(2+t^{\prime2}-t^2)N_n(t,t^\prime)$} which follows directly from the proof of DJ95P1 with  $N_n(t,t^\prime)=\sum_{i=1}^{n}\sigma_i^2 I(t<|Y_i\sigma_i^{-1}|\le t^{\prime})$. The second term is bounded by $n^{-1}(3+t^{\prime2})M_{n}(\tilde{\tau}_{k},\tilde{\tau}_{k}^\prime)$ where $M_{n}(\tilde{\tau},\tilde{\tau}^\prime)=\sum_{i=1}^{n}\sigma_i^2 I(\tilde{\tau}<F_{S_i}(s_i)<\tilde{\tau}^{\prime})$ and similarly the third term is bounded by $n^{-1}(3+t^{\prime2})M_{n}(\tilde{\tau}_{k-1},\tilde{\tau}_{k-1}^\prime)$.

For the risk $r_n^k(\tilde{\mfk{T}},\bm{\theta})$, we follow the same decomposition and upper bound $|r_n^k(\tilde{\mfk{T}},\bm{\theta})-r_n^k(\tilde{\mfk{T}}^\prime,\bm{\theta})|$ by
\begin{eqnarray}
 \Big|r_n^k(\tilde{\mfk{T}},\bm{\theta})-r_n^k(\tilde{\mfk{T}}^{(3)},\bm{\theta})\Big|+\Big|r_n^k(\tilde{\mfk{T}}^{(3)},\bm{\theta})-r_n^k(\tilde{\mfk{T}}^{(2,3)},\bm{\theta})\Big|+\Big|r_n^k(\tilde{\mfk{T}}^{(2,3)},\bm{\theta})-r_n^k(\tilde{\mfk{T}}^\prime,\bm{\theta})\Big|\nonumber
\end{eqnarray}
From the proof of DJ95P1, we upper bound the first term above by $5n^{-1}\delta_{0n}t_n\sum_{i=1}^{n}\sigma_i^2$ as long as $|t-t^\prime|<\delta_{0n}$ for some $\delta_{0n}>0$. The second and the third terms are upper-bounded by $2n^{-1}\delta_{1n}(1+t_n^2)\sum_{i=1}^{n}\sigma_i^2$ and $2n^{-1}\delta_{2n}(1+t_n^2)\sum_{i=1}^{n}\sigma_i^2$ respectively as long as $|\tilde{\tau}_k-\tilde{\tau}_k^\prime|<\delta_{1n}$ and $|\tilde{\tau}_{k-1}-\tilde{\tau}_{k-1}^\prime|<\delta_{2n}$ for some $\delta_{1n},\delta_{2n}>0$.

Hence, we can bound $n\Big|V_n(\tilde{\mfk{T}}_k)-V_n(\tilde{\mfk{T}}_k^\prime)\Big|$ by
\begin{eqnarray}
\label{eq:1a.2}
&&\Big(2+t^{\prime2}-t^2\Big)N_n(t,t^\prime)+\Big(3+t^{\prime2}\Big)\Bigl\{M_{n}(\tilde{\tau}_{k},\tilde{\tau}_{k}^\prime)+M_{n}(\tilde{\tau}_{k-1},\tilde{\tau}_{k-1}^\prime)\Bigr\}+\\
&& 5\delta_{0n}t_n\sum_{i=1}^{n}\sigma_i^2+2\Big(\delta_{1n}+\delta_{2n}\Big)\Big(1+t_n^2\Big)\sum_{i=1}^{n}\sigma_i^2\nonumber
\end{eqnarray}
Following the proof of DJ95P1, we choose $\delta_{0n},\delta_{1n},\delta_{2n}$ such that $\delta_{0n}t_n, \delta_{1n}t_n^2$ and $\delta_{2n}t_n^2$ are all $o(n^{-1/2})$ and for large $n$ we use $\mbb{E}~ N_n(t,t^\prime)+(3+t_n^2)\{\mbb{E}~M_n(\tilde{\tau}_{k},\tilde{\tau}_{k}^\prime)+\mbb{E}~M_{n}(\tilde{\tau}_{k-1},\tilde{\tau}_{k-1}^\prime)\}\le c_0n\delta_{0n}+c_1n\delta_{1n}+c_2n\delta_{2n}=O(r_n n^{1/2})$ for some absolute constants $c_0,c_1,c_2$. This and the bound in equation \eqref{eq:1a.2} establish $r_n/\sqrt{n}=O(c_n^{-1})$ directly from the proof of DJ95P1 which proves the desired $\ell_1$ convergence of equation \eqref{eq:1a.1}.
\medskip\\
\noindent\textbf{Proof of Theorem 1, statement (b)} - Due to the result of theorem 1 part (a), proving the result in part (b) essentially reduces to showing $c_n\;\mbb{E}\Big[\sup_{\mfk{T}\in\mathcal{H}_n}\Big|l_n\{\bs{\theta},\hat{\bm{\theta}}^{SI}(\mfk{T})\}-r_n(\mfk{T},\bm{\theta})\Big|\Big]<\infty$. Note that the loss $l_n\{\bs{\theta},\hat{\bm{\theta}}^{SI}(\mfk{T})\}$ decomposes as the sum of $K$ losses: $l_n\{\bs{\theta},\hat{\bm{\theta}}^{SI}(\mfk{T})\}=\sum_{k=1}^{K} l_n^k\{\bs{\theta},\hat{\bm{\theta}}^{SI}(\mfk{T}_k)\}$ where $l_n^k\{\bs{\theta},\hat{\bm{\theta}}^{SI}(\mfk{T}_k)\}=n^{-1}\sum_{i=1}^{n}\Bigl\{Y_i+\sigma_i\eta_{t_k}(Y_i)-\theta_i\Bigr\}^2I\Bigl\{S_i\in(\tau_{k-1},\tau_k]\Bigr\}$ and $\mfk{T}_k=\{\tau_{k-1},\tau_k,t_k\}$. As the risk is just the expectation of the loss, by triangle inequality, it is enough to show
\begin{equation*}
c_n\;\mbb{E}\Big[\sup_{\mfk{T}_k\in \mbf{R}^2\times [0,t_n]}\Big|l_n^k\{\bs{\theta},\hat{\bm{\theta}}^{SI}(\mfk{T}_k)\}-\mbb{E }~l_n^k(\bs{\theta},\hat{\bm{\theta}}^{SI}(\mfk{T}_k))\Big|\Big]<\infty\text{ for all }i\text{ and for all large }n	
\end{equation*}
Note that each of the losses $l_n^k$ again decomposes into two parts:
\begin{eqnarray}
A_n&=& n^{-1}\sum_{i=1}^{n}\Bigl\{\sigma_iZ_i-\sigma_it_k\text{sign}(\theta_i+\sigma_iZ_i)\Bigr\}^2I\Bigl\{S_i\in(\tau_{k-1},\tau_k]\Bigr\}I\Bigl\{|\theta_i+\sigma_iZ_i|>\sigma_it_k\Bigr\}\nonumber\\
B_n&=& n^{-1}\sum_{i=1}^{n}\theta_i^2I\Bigl\{S_i\in(\tau_{k-1},\tau_k]\Bigr\}I\Bigl\{|\theta_i+\sigma_iZ_i|\le\sigma_it_k\Bigr\}\nonumber
\end{eqnarray}
where $Z_i$'s are i.i.d $N(0,1)$ random variables.

We next prove that for some $\epsilon_0$, (i) there exist functions $\mathfrak{g}_n$ and $\mathfrak{h}_n$ such that $\mbb{P}\Bigl\{c_n\sup_{\mfk{T}_k}\Big|A_n-\mbb{E}~A_n\Big|>\epsilon\Bigr\}\le\mathfrak{g}_n(\epsilon)$ and $\mbb{P}\Bigl\{c_n\sup_{\mfk{T}_k}\Big|B_n-\mbb{E}~B_n\Big|>\epsilon\Bigr\}\le\mathfrak{h}_n(\epsilon)$ for all $\epsilon>\epsilon_0$ and for all large $n$, and (ii) both $\limsup\int_{\epsilon_0}^{\infty}\mathfrak{g}_n(\epsilon)d\epsilon<\infty$, $\limsup\int_{\epsilon_0}^{\infty}\mathfrak{h}_n(\epsilon)d\epsilon<\infty$. This establishes the desired result.

We deal with $B_n$ first and, without loss of generality, establish the bound for\newline $B_n=n^{-1}\sum_{i=1}^{n}\theta_i^2I\Bigl\{S_i\in(\tau_{k-1},\tau_k]\Bigr\}I\Bigl\{\theta_i+\sigma_iZ_i\le \sigma_it_k\Bigr\}$. Now
\begin{eqnarray}
\label{eq:1b.2}
\mbb{P}\Bigl\{c_n\sup_{\mfk{T}_k}\Big|B_n-\mbb{E}~B_n\Big|>\epsilon\Bigr\}=\mbb{P}\Bigl\{c_n\sup_{\mfk{T}_k}\Big|B_n-\mbb{E}~B_n\Big|>\epsilon\text{ and }\mc{F}_n\Bigr\}+\mbb{P}\Big(\mc{F}_n^c\Big)
\end{eqnarray}
where the set $\mc{F}_n=\{\max_{i=1,\ldots,n}|Z_i|\le (1+\epsilon)\sqrt{2\log n}\}$, and $\mbb{P}\Big(\mc{F}_n^c\Big)\le \phi(0)n^{-\epsilon}$ for all large $n$. We bound the first term on the right side of equation \eqref{eq:1b.2} by using the Glivenko-Cantelli theorem for weighted empirical measures \citep{singh1975glivenko}. As $t_n=\sqrt{2\log n}$ and $t_k\in[0,t_n]$, on $\mc{F}_n$ the weights in $B_n$ can be positive only when $\theta_i^2\le (2+\epsilon)^2\sigma_i^2~2\log n$. We next use the inequality in equation (6) of \cite{singh1975glivenko} with $a$ in that equation equaling $\epsilon c_n^{-1}\sum_{i=1}^{n}\sigma_i^4$. Further, note that for all large $n$, $\epsilon c_n^{-1}\sum_{i=1}^{n}\sigma_i^4\ge\sqrt{(2+\epsilon)^4~(2\log n)^2~\sum_{i=1}^{n}\sigma_i^4~}$ which is the maximum possible $\ell_2$ norm of the weights $\theta_i^2$. This, along with assumption A1 and the fact that $\sum_{i=1}^{n}\theta_i^2\le \sqrt{n}(\sum_{i=1}^{n}\theta_i^4)^{1/2}$ gives
\begin{eqnarray}
\mbb{P}\Bigl\{\sup_{\mfk{T}_k}c_n\Big|B_n-\mbb{E}~B_n\Big|>\epsilon,~\mc{F}_n\Bigr\}<4\dfrac{n\epsilon \log^{\delta}n}{\sqrt{\sum_{i=1}^{n}\theta_i^4}}\exp{\Bigl\{-\dfrac{\epsilon^2\log^{2(\delta-1)} n}{2(2+\epsilon)^4}\Bigr\}}\nonumber
\end{eqnarray} 
Now if $\sum_{i=1}^{n}\theta_i^4=o(c_n^{-1})$, then the desired bound on $\mbb{E}\sup_{\mfk{T}_k}c_n\Big|B_n-\mbb{E}~B_n\Big|$ is obvious; else the above probability is bounded above by $\mathfrak{h}_n(\epsilon)=4n^2\epsilon\exp{\Bigl\{-\epsilon^2\log ^{2(\delta-1)} n/2(2+\epsilon)^4\Bigr\}}$ which satisfies the aforementioned integrability condition. Thus, the proof of the result for $B_n$ is complete.

We now turn our attention to $A_n$. Again, without loss of generality, we prove the bound for $A_n=n^{-1}\sum_{i=1}^{n}(Z_i-t_k)^2I\Bigl\{S_i\in(\tau_{k-1},\tau_k]\Bigr\}I\Bigl\{\theta_i+Z_i>t_k\Bigr\}$. As we saw in the case of $B_n$, the variances $\sigma_i^2$ appear only through $n^{-1}\sum_{i=1}^{n}\sigma_i^2$ which is finite by assumption A1. Thus, we take $\sigma_i=1$ for all $i$ and decompose $A_n$ as sum of three parts by expanding $(Z_i-t_k)^2=Z_i^2-2Z_it_k+t_k^2$. The bound on the third term follows directly by the traditional Glivenko-Cantelli theorem and by noting that $t_k^2\le 2\log n$. Here we establish the $\ell_1$ convergence result for the first term. The proof for the second term is very similar.

We further reduce the problem. Without loss of generality, we assume $\theta_i=0$ and prove the $\ell_1$ convergence result for
$A_n=n^{-1}\sum_{i=1}^{n}Z_i^2I\Bigl\{S_i\in(\tau_{k-1},\tau_k]\Bigr\}I\Bigl\{Z_i>t_k\Bigr\}$. We again apply the same technique as with $B_n$ and control the probability $\mbb{P}\Bigl\{\sup_{\mfk{T}_k}c_n\Big|A_n-\mbb{E}~A_n\Big|>\epsilon,~\mc{F}_n\Bigr\}$. Similarly as with $B_n$, but now conditioned on $\{Z_i:i=1,\ldots,n\}$, the above probability is easily controlled at the desired rate by applying equation (6) of \cite{singh1975glivenko}, i.e,
$\mbb{P}\Bigl\{\sup_{\mfk{T}_k}c_n\Big|A_n-\mbb{E}~A_n\Big|>\epsilon,~\mc{F}_n\mid Z_1,\ldots,Z_n\Bigr\}\le \mathfrak{g}_n(\epsilon)$ where $\mathfrak{g}_n$ does not depend on $Z_i$ and for some $\epsilon_0>0$, $\int_{\epsilon_0}^{\infty}\mathfrak{g}_n(\epsilon)d\epsilon<\infty$ for all large $n$ with $\sum_{i=1}^{n}Z_i^2c_n\to\infty$ as $n\to\infty$.

This establishes the desired $\ell_1$ result for $A_n$ and completes the proof.
\medskip\\
\noindent \textbf{Proof of Corollary 1} - Both statements of this corollary follow from result (b) of Theorem 1. For statement (a), note that for any $\epsilon>0$, the probability  $\mbb{P}\Big[l_n\{\bm{\theta},\hat{\bm{\theta}}^{SI}(\hat{\mfk{T}})\}\ge l_n\{\bm{\theta},\hat{\bm{\theta}}^{SI}(\mfk{T}^{OL})\}+c_n^{-1}\epsilon\Big]$ is bounded above by $\mbb{P}\Big[l_n\{\bm{\theta},\hat{\bm{\theta}}^{SI}(\hat{\mfk{T}})\}-S(\hat{\mfk{T}},\pmb Y,\bm S)\ge l_n\{\bm{\theta},\hat{\bm{\theta}}^{SI}(\mfk{T}^{OL})\}-S(\mfk{T}^{OL},\pmb Y,\bm S)+c_n^{-1}\epsilon\Big]$, which converges to $0$ by Theorem 1 (b).

Statement (b) of this corollary follows  as the difference $l_n\{\bm{\theta},\hat{\bm{\theta}}^{SI}(\hat{\mfk{T}})\}- l_n\{\bm{\theta},\hat{\bm{\theta}}^{SI}(\mfk{T}^{OL})\}$ can be decomposed as sum of $l_n\{\bm{\theta},\hat{\bm{\theta}}^{SI}(\hat{\mfk{T}})\}-S(\hat{\mfk{T}},\pmb Y,\bm S)$, $S(\mfk{T}^{OL},\pmb Y,\bm S)-l_n\{\bm{\theta},\hat{\bm{\theta}}^{SI}(\mfk{T}^{OL})\}$ and
$S(\hat{\mfk{T}}\pmb, Y,\bm S)-S(\mfk{T}^{OL},\pmb Y,\bm S)$. By definition, the last term is not positive. Thus, the sum is bounded above by\newline
$ 2\sup_{\mfk{T}\in\mathcal{H}_n} |S(\mfk{T},\pmb Y,\bm S)-l_n\{\bs{\theta},\hat{\bm{\theta}}^{SI}(\mfk{T})\}|$ which converges to $0$ at the prescribed rate by Theorem 1 (b).
\section{Detailed proofs of the results of section 3 of the main paper}
\label{supp:moreproofs_3}
We use  Theorem 2 of \citet{johnstone1994minimax}, which provides an explicit higher order evaluation of the maximal risk of the soft threshold estimator with the best possible choice of threshold. We restate the theorem abet for the symmetric case, which slightly increases the maximal risk presented in equation (17) of the aforementioned theorem. 
\begin{result}
	Consider the class of univariate soft-threshold estimators $\hat \theta^S_{\lambda}(x)=\text{sign}(x)(x-\lambda)_+$ for $\lambda \geq 0$. If the parameter $\theta \in \mbf{R}$ is such that $P(\theta=0)\leq 1-\eta$. Then as $\eta \to 0$, the best choice of threshold is $f(t)+O(t^{-3}\log t)$ and the minimal possible risk is $H(t)=\eta (h(t)+36\,t^{-2}\log t+O(t^{-2}))$ where $t=\sqrt{2 \log \eta^{-1}}$ and 
	\begin{align}
	\label{def.opt.thr}
	f(t)=\sqrt{t^2-6\log t+ 2 \log\phi(0)} \text{ and } h(t)=f^2(t)+5~.
	\end{align}
\end{result} 
\noindent\textbf{Proof of Theorem 2} - 
Directly applying the above result we have,
\begin{align*}
&\mc{R}^{NS}_n(\alpha,\beta,\rho_n) =(\pi_{1,n} n^{-\alpha}+\pi_{2,n} n^{-\beta})H(t_{n}) \, \bar{\sigma^2_n} , \text{where, }  \bar{\sigma^2_n}=n^{-1}\sum_{i=1}^{n}\sigma_i^2~, 
\end{align*}
and $t^2_{n}=2\log(\pi_{1,n} n^{-\alpha}+\pi_{2,n} n^{-\beta})^{-1}$ as the density level  is at most $\pi_{1,n} n^{-\alpha}+\pi_{2,n} n^{-\beta}$ in $\Theta_n(\alpha,\beta,\rho_n)$. 
Now, if we completely know the latent side information then again applying equation (17) of \cite{johnstone1994minimax} separately to the two groups: $\{i: \xi_i \le \tau_n^\star\}$ and $\{i: \xi_i> \tau_n^\star\}$ we have:
\begin{align*}\mc{R}^{OS}_n(\alpha,\beta,\rho_n)= \{\pi_{1,n} n^{-\alpha}H(t_{1,n})+\pi_{2,n} n^{-\beta}H(t_{2,n})\}\,\bar{\sigma^2_n} \text{ where } t^{2}_{1,n}=2\alpha k_n, t^{2}_{2,n}=2\beta k_n.
\end{align*}
Also, $t_{n}^2=t_{1,n}^2-2\log \pi_{1,n}+O(\pi_{2,n} \pi_{1,n}^{-1}n^{\alpha-\beta})$. By Assumption (A2.1) there exists $\epsilon > 0$ such that $\pi_{2,n} \pi_{1,n}^{-1}n^{\alpha-\beta} < n^{-\epsilon}$ for all large $n$. Thus, as $n \to \infty$ with $c_0=5+2 \log \phi(0)$, and $\tilde{k}_n=k_n/\log k_n$,
\begin{align*}
\mc{R}^{OS}_n&= \pi_{1,n}\, p_{1,n}\, \bar{\sigma^2_n}\,\{2 \alpha k_n-3 \log (2\alpha k_n) + c_0+O(\kt)\}\\
\mc{R}^{NS}_n&= \pi_{1,n}\, p_{1,n}\, \bar{\sigma^2_n}\,\big[2 \alpha k_n+ 2 \log \pi_{1,n}^{-1}-3 \log (2\alpha k_n) - 3 \log\big\{1+(\alpha k_n)^{-1}\log \pi_{1,n}^{-1}\big\}+ c_0+O(\kt)\big]~,
\end{align*}
from which the  lemma follows. 

To understand the phenomenon here in a simplifier lens, consider the first order  approximations: $\mc{R}^{NS}_n\sim \bar{\sigma^2_n}\pi_{1,n} p_{1,n}f(t_{n})$, $\mc{R}^{OS}_n\sim \bar{\sigma^2_n}\pi_{1,n} p_{1,n} f(t_{1,n})$, $f(t_{n})-f(t_{1,n})\sim 2\log\pi_{1,n}^{-1}$ and $$\mc{R}^{NS}_n-\mc{R}^{OS}_n\sim (2 \log \pi_{1,n}^{-1})\,\pi_{1,n}\, p_{1,n}\, \bar{\sigma^2_n}$$
Thus, the gain due to incorporation of side information is essentially due to the fact that we can use a lower threshold for the subgroup with smaller sparsity than that used in the agglomerative case with no side information and this is exactly the phenomenon depicted in Figure 2 of the main paper.
\medskip\\
\noindent\textbf{Proofs of Theorems 3, 4 and Lemma 2} - Note that in our asymptotic set-up, there exists $\epsilon > 0$ such that 
\begin{align}\label{eq.app1}
p_{1,n}\pi_{1,n}(p_{2,n}\pi_{2,n})^{-1} \geq n^{\epsilon} \text{ for all large } n.
\end{align}
We will be using this property to simplify our calculations by restricting ourselves to dominant terms. As such we will be ignoring terms which are $o(p_{1,n}\pi_{1,n}\bar{\sigma_n^2}k_n^{-1})$.  Without loss of generality we assume that ${S}|\xi$ has monotone likelihood ratio in ${S}$ and consider $ q^{jk}_{i,n}(\tau)\coloneqq
\mbb{P}_n(\hat I_i^j|I_i^k\Big) \text{ for } j, k\in\{1,2\}, \;i\in \{1,\ldots,n\},
$
where,  $\hat I_i^1=\{S_i\le {\tau}\}$, $I_i^1=\{\xi_i\le \tau^\star_n\}$, $\hat I_i^2=\mbf{R}\setminus \hat I_i^1$, $\I_i^2=\mbf{R}\setminus I_i^1$ and $q^{jk}_{n}(\tau)=n \sum_{i=1}^nq^{jk}_{i,n}(\tau) \sigma_i^2/\bar{\sigma_n^2}$.
\medskip\\
\noindent \textbf{Proof of Lemma~2 - }
Dividing the difference in Theorem~2 by the expression of $\mc{R}_n^{OS}$ from the display above it we get
\begin{align}\label{eq.temp.11}
\mc{R}^{NS}_n/\mc{R}^{OS}_n = 1+ [2 \log \pi_{1,n}^{-1}- 3 \log\{1+(\alpha k_n)^{-1}\log \pi_{1,n}^{-1}\}]\big/\{2 \alpha k_n-3\log (2\alpha k_n)+c_0\}+O(k_n^{-\nu}),
\end{align}
where $c_0=5+2 \log \phi(0)$ and $\nu<2$. The first result of the lemma now follows by noting  $\log \pi_{1,n}^{-1}\leq \gamma_0 < \beta -\alpha$ which is due to Assumption (A2.1).
\par
Note that, iff $c_n \to 0$ then $\mc{R}^{NS}_n/\mc{R}^{OS}_n \to 1$. From the above display it follows that 
$k_n^{1+\delta}(\mc{R}^{NS}_n/\mc{R}^{OS}_n -1) \geq k_n^{\delta} \log \pi_{1,n}^{-1} \{1-1.5(\alpha k_n)^{-1}\} +O(k_n^{-\nu+1+\delta})$ where $\nu <2$. {Thus, we have the first part of the third result. Its second part  follows directly from the proof of Theorem~3, which is present after the proof of this lemma}.\\
Next, we establish the upper bound on the maximal risk of ASUS given in the second statement of lemma 2. Let $\mc{R}_n^{KS}$  denotes the maximal risk of ASUS when we can set any possible thresholds in ASUS including those depending on the density levels $p_{1,n}, p_{2,n}$ as well as the mixing probabilities $\pi_{1,n}$ and $\pi_{2,n}$. However, we do not know the latent variable or its subsequent oracle optimal groups $\{i: \xi_i \leq \tau^*_n\}$ and $\{i: \xi_i > \tau^*_n\}$.
Thus, by definition $\mc{R}_n^{OS} \leq \mc{R}_n^{KS}\leq\mc{R}_n^{NS}$.  Now, ASUS always chooses the thresholds and the segmentation hyper-parameter in a data-dependent fashion minimizing the SURE criterion. We next apply theorem~1 which tells us that the maximal risk of ASUS $\mc{R}_n^{AS}$ can not be much bigger than $\mc{R}_n^{KS}$.
As such, theorem~1 compounded with theorem~4a of DJ95 implies  $\mc{R}_n^{AS}-\mc{R}_n^{KS} \leq \mc{R}_n^{F}I\{\mu_n^2 \leq 3 d_n\}+o(\pi_{1,n}p_{1,n}\bar{\sigma_n^2}k_n^{-1})$ where $\mc{R}_n^{F}$ is the risk of ASUS with fixed threshold of $\sqrt{2 k_n}$ and $d_n=n^{-1/2}\log^{3/2} n$ and $\mu^2_n=n^{-1} \sum_{i=1}^n \theta_i^2 \wedge (2 k_n)$. By Lemma~8.3 of \cite{johnstonebook}, 
$\mc{R}_n^{F}\leq n^{-1}+n^{-1} \sum_{i=1}^n \{\theta_i^2 \wedge (1 + 2 k_n)\} \sigma_i^2$.  
Thus, $\mc{R}_n^{AS}\leq \mc{R}_n^{NS}+o(\pi_{1,n}p_{1,n}\bar{\sigma_n^2}k_n^{-1})$ and the result follows from \eqref{eq.temp.11}.
\medskip\\
	\noindent \textbf{Proof of Theorem 3 - }First consider the situation where the sparsity levels $p_{1,n}$ and $p_{2,n}$ are known. 
	Due to the product structure of our ASUS estimator, we first concentrate on its maximal risk for each of the $i$th coordinate. This reduces to an univariate risk analysis. If noise variance equals 1, univariate soft-threshold estimators with threshold $\lambda$ has:\\
	(a) the risk at the origin: $g_1(\lambda)(1+O(\lambda^{-2}))$ for large $\lambda$ where $g_1(\lambda)=4 \phi(\lambda)/\lambda^3$ \\
	(b) the maximal risk at the non-origin points: $g_2(\lambda)=1+\lambda^2$ and the maximum is attained when the parametric value is $\pm \infty$. \\
	Now, if the probability of the parameter $\theta_i$ being non-zero is $p$ then the maximal risk of the soft-threshold estimator with threshold $\lambda$ is $g(p,\lambda)=(1-p)g_1(\lambda)+p g_2(\lambda)$. 
	
\begin{figure}[t]
		\centering
		\includegraphics[width=1\linewidth]{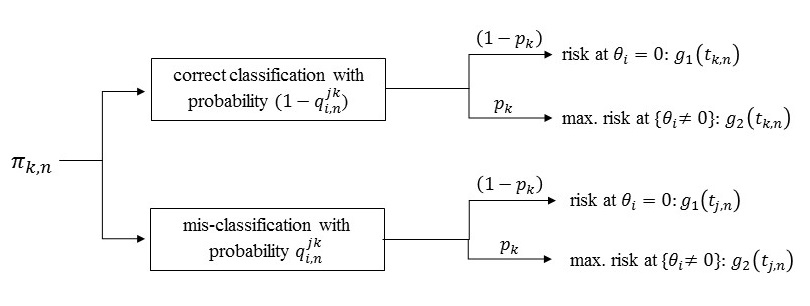}
		\caption{Pictorial representation of coordinate-wise decomposition of maximal risk $\mc{R}^{AS}$. Here $j,k=1,2$ and $j\ne k$.}
		\label{fig3.2}
	\end{figure}

	As $\theta_i$ is generated from the two group model of equation (8) of the main paper with density levels $p_{k,n}$, 
	$q^{12}_{i,n}$ and $q^{21}_{i,n}$ are the probabilities of mis-classifying group 2 and group 1 respectively and thresholds $t_{k,n}$ were used for those detected in group $k=1,2$. Note, that without mis-classification the maximal risk at each coordinate $i$ is weighted by the group probabilities $\pi_{k,n}$ and the optimal threshold choices are $f(\sqrt{2\alpha k_n})$ and $f(\sqrt{2\beta k_n})$ where,
	$f$ is defined in \eqref{def.opt.thr}.
	However, under mis-classification these thresholds will change and the optimal thresholds will be $f(m^{\sf opt}_{1,n}[i])$ and $f(m^{\sf opt}_{2,n}[i])$ where, 
	\begin{align}\label{eq.app6}
	m_{k,n}^{\sf opt}[i]=\bigg\{-2\log\bigg(\frac{\pi_{k,n}\,q_{i,n}^{kk}\,p_{k,n}+\pi_{j,n}\,q_{i,n}^{kj}\,p_{j,n}}{\pi_{k,n}\,q_{i,n}^{kk}+\pi_{j,n}\,q_{i,n}^{kj}}\bigg)\bigg\}^{1/2} \text{ for } j\neq k~.
	\end{align}
	These can not be used as $q^{jk}_{i,n}$ are not known while constructing the estimator. 
	We are interesting in deriving upper bounds on the maximal risk of ASUS and so, unlike the optimal thresholds which depend on $i$, here we consider thresholds $t_{1,n}$ and $t_{2,n}$ which are uniform over the groups.
	With mis-classification, the probabilities $q^{jk}_{i,n}$ will be also involved into the expression for maximal risk as now $\theta_i$ coming from group $1$ (say) might be treated with either threshold $t_{1,n}$ (when correctly classified) or  $t_{2,n}$ (when incorrectly classified). Figure \ref{fig3.2} provides a pictorial representation of how the probabilities $q^{jk}_{i,n}$ enter this decomposition.
	The maximal risk for coordinate $i$ is given by
	\begin{equation}
	\label{p2.1}
	\sum_{j=1}^{2}\sum_{k=1}^{2}\pi_{k,n}\,q^{jk}_{i,n}\,g(p_{k,n},t_{j,n}) \sigma_i^2 \{1+o(1)\}~. 
	\end{equation}
	We fix a threshold of $t_{1,n}=f(\sqrt{2\alpha k_n})$ and $t_{2,n}=f(\sqrt{2 \gamma k_n})$ where $\gamma$ is allowed to vary in $[\alpha,\beta]$. These thresholds when substituted in \eqref{p2.1} will produce a upper bound on the maximal risk. Doing so and using \eqref{eq.app1}, we see that the maximal risk for the $i$th coordinate is upper bounded by 
	\begin{equation}
	\label{p2.2}
	\sigma_i^2\Big[\pi_{1,n} p_{1,n}\Big\{q_{i,n}^{11}g_2(t_{1,n})+q_{i,n}^{21}g_2(t_{2,n})\Big\}+\sum_{j=1}^{2}\sum_{k=1}^{2}\pi_{k,n}\,q_{i,n}^{jk}\,g_1(t_{j,n})+O\Big(\dfrac{\pi_{1,n}\,p_{1,n}}{k_n}\Big)\Big]
	\end{equation}
	Now, consider the second term in equation \eqref{p2.2}. As $t_{2,n} \geq t_{1,n}$, we lower bound  
	and upper bound it by $A_n+\pi_{1,n}q_{i,n}^{11}g_{1}(t_{1,n})$ and $A_n+\pi_{1,n}g_{1}(t_{1,n})$
	where $A_n=\pi_{2,n}q_{i,n}^{22}g_1(t_{2,n})+\pi_{2,n}q_{i,n}^{12}g_{1}(t_{1,n})$. \\
	Define $\tilde k_n=k_n/\log k_n$. 
	Note that $g_1(t_{1,n})=4 p_{1,n}\{1+O(\kt)\}$ and $g_1(t_{2,n})=4 \phi(0)t_{2,n}^{-3}\phi(t_{2,n})(1+O(k_n^{-1}))$. Thus, with $\kappa_n(\gamma)=n^{\alpha}\phi(0) \phi(t_{2,n})t_{2,n}^{-3}$,\, $t_{2,n}=f(\sqrt{2\gamma k_n})$ the second term in \eqref{p2.2} is bounded above by
	\begin{align}\label{p2.3}
	{4}\pi_{2,n}p_{2,n}\{{ q_{i,n}^{22}\rho_n}\kappa_n(\gamma)+{\rho_n q_{i,n}^{12}}+1\}\{1+O(\kt)\}
	\end{align}
	as $\rho_n=\pi_{2,n}/\pi_{1,n}$.
	Now, consider the first term in equation \eqref{p2.2}. We have,
	\begin{eqnarray}
	q_{i,n}^{11}\,g_2(t_{1,n})+q_{i,n}^{21}\,g_2(t_{2,n})=g_2(t_{1,n})+q_{i,n}^{21}\{g_2(t_{2,n})-g_2(t_{1,n})\}\nonumber
	\end{eqnarray}
	and $g_2(t_{2,n})-g_2(t_{1,n})=2(\gamma-\alpha)k_n-3 \log \log (\gamma/\alpha):=\delta_n(\gamma)$. 
	Thus, the first term in equation \eqref{p2.2},
	\begin{equation}
	\label{p2.4}
	\pi_{1,n} p_{1,n}\Bigl\{q_{i,n}^{11}g_2(t_{1,n})+q_{i,n}^{21}g_2(t_{2,n})\Bigr\}=\pi_{1,n} p_{1,n}\Bigl\{g_2(t_{1,n})+ q_{i,n}^{21}\,\delta_n(\gamma)\Big\}
	\end{equation}
	Now, $g_2(t_{1,n})=1+t_{1,n}^2=h(\sqrt{2\alpha k_n})-4$, and so, from equations \eqref{p2.3} and \eqref{p2.4}, maximal risk of ASUS for coordinate $i$ is upper bounded by
	\begin{equation*}
	\pi_{1,n}p_{1,n}\sigma_i^2\Big[h(\sqrt{2\alpha k_n})+\{\,q_{i,n}^{21}\,\delta_n(\gamma)+{4} {\rho_n\,q_{i,n}^{12}}+4 q_{i,n}^{22}\rho_n  \kappa_n(\gamma)\}\{1+O(\kt)\} +O\Big(\kt\Big)\Big]
	\end{equation*}
	and therefore the maximal risk over the $n$ coordinates of the ASUS estimator when thresholds can be directly chosen depending on the density levels $p_{1,n}$ and $p_{2,n}$  is
	\begin{equation*}
	\mc{R}_n^{KS}\le \pi_{1,n}p_{1,n} \bar{\sigma_n^2}\Big[h(\sqrt{2\alpha k_n})+\{\,q_{n}^{21}\,\delta_n(\gamma)+{4} {\rho_n\,q_{n}^{12}}+4 q_{n}^{22}\rho_n  \kappa_n(\gamma)\}\{1+O(\kt)\} +O\Big(\kt\Big)\Big]
	\end{equation*}
	where $q^{jk}_n(\tau)=\sum_{i=1}^n q^{jk}_{i,n}(\tau) \sigma_i^2\big/\sum_{i=1}^n \sigma_i^2$ for $j, k\in\{1,2\}$. Recall, $\mc{R}_n^{KS}$  denotes the maximal risk of ASUS when we can set any possible thresholds in ASUS including those depending on the density levels $p_{1,n}, p_{2,n}$ as well as the mixing probabilities $\pi_{1,n}$ and $\pi_{2,n}$. Now, consider the general case when those are unknown. ASUS always chooses the thresholds $t_{1,n}$ and $t_{2,n}$ and the segmentation hyper-parameter $\tau_n$ in a data-dependent fashion minimizing the SURE criterion. We next apply theorem~1 similarly as in the proof of lemma~2 which provides us with  $\mc{R}_n^{AS}-\mc{R}_n^{KS} \leq o(\pi_{1,n}p_{1,n}\bar{\sigma_n^2}k_n^{-1})$	
	Also, from calculation in the previous subsection for any $\nu < 1$, we know  $\mc{R}^{OS}_n\ge \pi_{1,n}p_{1,n} \bar{\sigma_n^2}\{h(\sqrt{2\alpha k_n})+o(k_n^{-\nu})\}$ and so
	\begin{equation}
	\label{eq.app2}
	\mc{R}^{AS}_n-\mc{R}^{OS}_n\le \pi_{1,n}p_{1,n}\bar{\sigma_n^2}\Big[\{\,q_{n}^{21}\,\delta_n(\gamma)+{4} {\rho_n\,q_{n}^{12}}+4 q_{n}^{22}\rho_n  \kappa_n(\gamma)\}\{1+O(\kt)\} +O(\kt)\Big]~.
	\end{equation}
	Again, in our asymptotic set-up there exists $\epsilon >0$ such that $\rho_n  \leq n^{\gamma_0-\epsilon}$ for all large $n$.  Choosing $\gamma=\alpha+\gamma_0$,we have $\delta_n(\gamma)=2\gamma_0 k_n-3 \log \log (1+\gamma_0/\alpha)$ and $\rho_n \kappa_n= o(n^{-\epsilon'})$ for some  $0 < \epsilon' < \epsilon$. Thus, with this choice of $\gamma$, based on  \eqref{eq.app2} the controls on $q^{12}_n$ and $q^{21}_n$ stated in the theorem implies $\mc{R}^{AS}_n-\mc{R}^{OS}_n\le O(\kt)$. This, along with theorem~2 provide us with the desired result for the theorem as well as the third result of lemma~2.
	\medskip\\
\noindent {\textbf{Proof of Theorem 4, statement (a)} - }Consider case (ii) first. Note that $\lim_{n \to \infty} n \rho_n q^{21}_n(\tau_n) =0$ implies $\rho_nq_{i,n}^{21}(\tau_n)=o(1)$ as $n \to \infty$ for all $i$. Hence for each coordinate $i$, the optimal threshold for group 1 considering two groups in the data is $f(m_{1,n}^{\sf opt}[i])$ (see equation \eqref{eq.app6}). As  $\rho_nq_{i,n}^{21}(\tau_n)=o(1)$ , we have $m_{1,n}^{\sf opt}=\sqrt{2 \alpha k_n}\{1+o(k_n^{1-\epsilon})\}$ for any $\epsilon >0$. Thus, the threshold here asymptotically equals $t_{1,n}$ used in the proof of theorem~3: part b, before. Concentrating on only the $j=1, k=1$ and $j=2, k= 1$ terms in equation \eqref{p2.1} by the previously conducted analysis  we have the maximal risk of the $i$th coordinate bounded below by
$\pi_{1,n}p_{1,n} \sigma_i^2\Big[ h(\sqrt{2\alpha k_n})+{4 \rho_n\,q_{i,n}^{12}} \{1 +O(\kt)\}\Big]$.
Thus, if ASUS considers two groups then 
$\mc{R}^{KS}_n\geq \pi_{1,n}p_{1,n} \bar{\sigma_n^2}[ h(\sqrt{2\alpha k_n})+{4 \rho_n\,q_{i,n}^{12}} \{1 +O(\kt)\}]$
and the ratio  $(\mc{R}^{NS}_n-\mc{R}^{OS}_n)/(\mc{R}^{KS}_n-\mc{R}^{OS}_n)$ diverges to $\infty$ unless  $ \limsup \rho_n\,q_{n}^{12}k_n^{-1} < \infty$ as $\log \pi_{1,n}^{-1}=\gamma_0 k_n$.  In that case we use a uniform choice of threshold $t_{1,n}=t_{2,n}=f(m^{\sf opt}_n)$ where $m^{\sf opt}_n=\{-2\log(\pi_{1,n}p_{1,n}+\pi_{2,n}p_{2,n})\}^{1/2}$.
This along with part (a) completes the proof for case (ii).

For case (i), note that the $m_{1,n}^{\sf opt}$ for the $i$th coordinate defined in \eqref{eq.app6} equals 
$$m_{1,n}^{\sf opt}[i]=\{2\log p_{1,n}^{-1}+2\log(1+\rho_n q^{21}_{i,n}/q^{11}_{i,n})\}^{1/2}\{1+o(1)\} \text{ as } n \to \infty~.$$
Now, considering the risk at the origin for $j=1, k=1$ term in equation \eqref{p2.1}, we see that it will contain at least an extra additive component of $O(\pi_{1,n} p_{1,n}\rho_n q^{21}_{i,n}/q^{11}_{i,n} \sigma_i^2)$ over $h(\sqrt{2 \alpha k_n})$.  Thus, the average maximal risk over the $n$ coordinates is bounded below by  $$\pi_{1,n} p_{1,n}\Bigl\{h(\sqrt{2\alpha k_n}) \bar{\sigma_n^2} + O\bigg( \rho_n n^{-1}\sum_{i=1}^n q^{21}_{i,n}/q^{11}_{i,n} \sigma_i^2 \bigg)\Bigr\}$$
As $q^{11}_{i,n} \leq 1$ for all $i$, the second term on right side above is bounded below by $O(\pi_{1,n} p_{1,n}\rho_n q^{21}_{n} \bar{\sigma_n^2})$ which provides us with the desired result.
\medskip\\
\noindent {\textbf{Proof of Theorem 4, statement (b)} - }By definition, {$\mc{R}^{KS}_n\leq \mc{R}^{NS}_n$} and by application of theorem~1 (as in the proof of lemma~2) we have $\mc{R}_n^{AS}-\mc{R}_n^{KS} \leq o(\pi_{1,n}p_{1,n}\bar{\sigma_n^2}k_n^{-1})$. 
Also, by assumption A2.2, there exists some $\nu <1$ such that $\lim k_n^{\nu/2}(1-\pi_1)=\infty$. This implies $ k_n^{\nu}\, \log \pi_{1,n}^{-2}$, which coupled with theorem~2 gives us with the desired result. 
\medskip\\
\noindent\textbf{Proof of Lemma~1 - }Without loss of generality, assume that the marginal distribution of the auxiliary sequence ${S}$ given the latent parameter $\xi$ has monotone likelihood ratio  in the statistic ${S}$. Also, let $\underline{\mu}_{2,n}=\inf\{\mu_i: i \in I^*_{2,n}\}>\sup\{\mu_i: i \in I^*_{1,n}\}=\bar{\mu}_{1,n}$ . Then, $\underline{\mu}_{2,n}=\bar{\mu}_{1,n}+d_n$.  Under this reduction $ q^{jk}_{i,n}(\tau)\coloneqq
\mbb{P}_n(\hat I_i^j|I_i^k\Big) \text{ for } j, k\in\{1,2\}, \;i\in \{1,\ldots,n\},
$
where, the sets $\hat I_i^1=\{S_i\le {\tau}\}$, $I_i^1=\{\xi_i\le \tau^\star_n\}$, $\hat I_i^2=\mbf{R}\setminus \hat I_i^1$, $\I_i^2=\mbf{R}\setminus I_i^1$. Let $\bar{b}_n=\sup_i \max(2\nu_i^{2},b_i)$.
Now, set $\tau_n=\bar{\mu}_{1,n}+\bar{b}_n^{\gamma}(2  k_n + \log k_n)^{\gamma}$. Thus, for all large $n$ the condition imposed on $d_n$ in the lemma implies {$\tau_n\leq\underline{\mu}_{2,n}-\bar b_n^{\gamma}(\log \rho_n+ \log k_n)^{\gamma}$}. Next, note that, 
$q^{21}_{n}(\tau_n)\le \bar{\sigma_n^2} \, \sup_{i} q^{21}_{i,n}(\tau_n) \leq \bar{\sigma_n^2} \, \sup_{i} P(S_i \leq \tau_n|\bar{\mu}_{1,n})$
and thereafter, using the tail bounds of the sub-Exponential/Gaussian distributions, we have for all $i$,
\begin{eqnarray*}
	P(S_i \leq \tau_n|\bar{\mu}_{1,n})  \leq \exp\{-{(\tau_n-\bar{\mu}_{1,n})^{1+I(b_i=0)}}/\max(2\nu_i^{2},b_i)\}\leq k_n^{-2}/\log k_n
\end{eqnarray*}
for all large n. Similarly, it follows that $q^{12}_{n}(\tau_n)\leq \rho_n^{-1}/\log k_n$, which completes the proof. 
\medskip\\		
\noindent {\textbf{Statement and proof of Lemma 3 - }}

\noindent\textbf{Lemma 3} - \textit{If our parametric space $\Theta(\alpha,\beta,\rho_n)$ is such that  $\limsup_{n \to \infty} n^{1/2}\, \pi_{1,n} p_{1,n} < \infty$, ASUS convergences in probability to the SureShrink procedure with the fixed threshold choice of $\sqrt{2 \log n}$.} 
\medskip\\
\noindent \textbf{Proof.} Define, $\mu_n^2=n^{-1}\sum_{i=1}^n \theta_i^2 \wedge (5 k_n)$ for some prefixed $\epsilon >0$. Note that $\mu_n^2 \leq 5 \pi_{1,n} p_{1,n} k_n (1+o(1))$ and so, by the condition of the lemma $\mu_n^2/d_n \to 0$ where $d_n=n^{-1/2} \log^{3/2} n$. 
Define $s_n^2=n^{-1}\sum_{i=1}^{n} y_i^2 \wedge (2k_n) -1$ where $Y_i=\theta_i +Z_i$ and $Z_i$ are i.i.d. $N(0,1)$ . 
Let $F_n=\{\max_i z_i^2< 3 k_n\}$. Note that $P(s_n^2 \leq d_n)\leq  P(F_n^c) + P(s_n^2 \leq d_n \text{ and } F_n).$ 
The firm term converge to $0$ and the second term on the right side above is bounded by $P(n^{-1}\sum_{i=1}^{n} Y_i^2 -1 \leq d_n, n^{-1}\sum_{i=1}^n\theta_i^2 \leq \mu_n^2)$ which converges to $0$ as $\mu_n^2/d_n \to 0$ (see proof of theorem 4 (b) of DJ95). This completes the proof. 	
\section{Additional simulation experiments}
\label{supp:moresims}
In this section, we present a number of simulation experiments demonstrating the asymptotic performance of ASUS as $n$ increases. We fix $m=50$ and allow $n$ to vary from $500$ to $5000$ in increments of $100$. To simulate the parameter vector $\bm{\theta}$, we continue to use the setup of the one sample problem discussed in Section 4.2.1 of the main paper and simulate $\eta_{1i}$ as before but vary the sparsity levels in $\bm{\theta}$ under scenarios S1 and S2 as follows:
\begin{itemize}
	\item[(S1)]$\bm{\xi}\sim\Big(\underbrace{\mathrm{Unif}(6,7)}_\text{ $1\%$ of $n$},\underbrace{\mathrm{Unif}(2,3)}_\text{$4\%$ of $n$},\underbrace{0,\ldots\ldots,0}_\text{$n-5\%$ of $n$}\Big)$
	\item[(S2)]$\bm{\xi}\sim\Big(\underbrace{\mathrm{Unif}(4,8)}_\text{$4\%$ of $n$},\underbrace{\mathrm{Unif}(1,3)}_\text{$16\%$ of $n$},\underbrace{0,\ldots\ldots,0}_\text{$n-20\%$ of $n$}\Big)$
\end{itemize}
with $\bm{\theta}=\bm{\xi}+\bm{\eta}_{1}$, $Y_i\sim N(\theta_i,\sigma_i^2)$ where $\sigma_i^2=1$ for all $i$ under S1 and $\sigma_i^2\stackrel{i.i.d}{\sim}\mathrm{Unif}(0.1,1)$ under S2. For scenario S1, we consider two side information $\bm S_1$ and $\bm S_2$ as follows: (ASUS.1) ${S}_{i1}|i\in\mc{I}_1^{\tau^\star}= |N(\mu_0,\sigma_i^2)+\bar{\eta}_{2i}|$, ${S}_{i1}|i\in\mc{I}_2^{\tau^\star}= |N(\mu_1,\sigma_i^2)+\bar{\eta}_{2i}|$ with $\mu_0=\sqrt{\log k_n}$, $\mu_1=0$, and (ASUS.2) ${S}_{i2}|i\in\mc{I}_1^{\tau^\star}= |N(\mu_0,\sigma_i^2)+\bar{\eta}_{2i}|$, ${S}_{i2}|i\in\mc{I}_2^{\tau^\star}= |N(\mu_1,\sigma_i^2)+\bar{\eta}_{2i}|$ with $\mu_0=\sqrt{k_n}$, $\mu_1=0$. Here $\sigma_i^2\stackrel{i.i.d}{\sim}\mathrm{Unif}(0.1,1)$ and $\bar{\eta}_{2i}$ is the average over $m$ samples of $N(0,0.01)$. Similarly for scenario S2, we consider two side information $\bm S_1$ and $\bm S_2$ as follows: (ASUS.1) ${S}_{i1}|i\in\mc{I}_1^{\tau^\star}=|\chi^2_{1+\sqrt{\log k_n}}+\bar{\eta}_{2i}|$, ${S}_{i1}|i\in\mc{I}_2^{\tau^\star}= |\chi^2_{1}+\bar{\eta}_{2i}|$, and (ASUS.2) ${S}_{i2}|i\in\mc{I}_1^{\tau^\star}= |\chi^2_{1+k_n}+\bar{\eta}_{2i}|$, ${S}_{i2}|i\in\mc{I}_2^{\tau^\star}= |\chi^2_{1}+\bar{\eta}_{2i}|$. Thus $\bm S_1$ and $\bm S_2$ differ in the separation of the means of their conditional distributions with $\bm S_2$ in scenario S2 having a near optimal separation in the means as prescribed by Lemma 1 in Section 3.3. 

\begin{figure}[!t]
	\centering
	\includegraphics[width=1\linewidth]{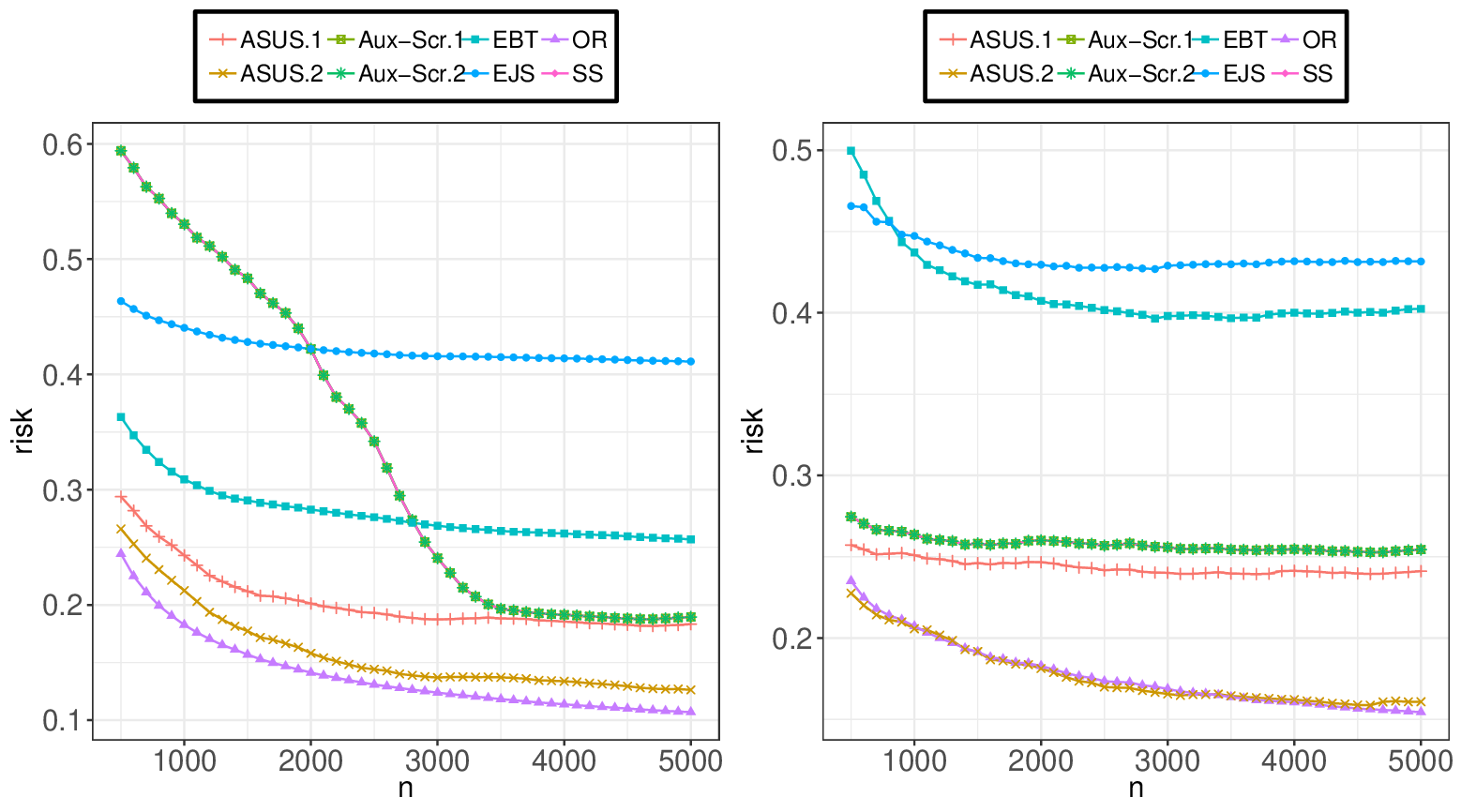}
	\caption{Asymptotic performance of ASUS: Average risks of different estimators. Dashed line represents the risk of the oracle estimator $\tilde{\theta}^{SI}_{i}(\mfk{T}^{OR})$. Left: Scenario S1 and Right: Scenario S2.}
	\label{simexp3}
\end{figure}

We repeat this sampling scheme for $N=1000$ repetitions and report the results in table \ref{tab:simexp3} and figure \ref{simexp3}. 
As observed in the one sample estimation problem, ASUS continues to exhibit the best performance amongst all the competing estimators. However, the efficiency of ASUS in exploiting side information clearly depends on the magnitude of separation of the conditional means of $\bm S$ under the two groups. For example ASUS.2, which has a bigger separation between the conditional means of side information $\bm S_2$, exhibits the best performance across all scenarios. In fact under scenario S2, $\bm S_2$ is sub-exponential and the $\log n$ separation between the conditional means brings the risk of ASUS.2 closer to the risk of the oracle estimator. As far as ASUS.1 is concerned, the relatively smaller separation in the conditional means of $\bm S_1$ does not allow ASUS to optimally partition the $n$ coordinates into heterogeneous groups in terms of sparsity and hence it performs only marginally better than the SureShrink estimator.  Moreover, we observe that 
\begin{table}[!t]
	\centering
	\caption{Asymptotic performance of ASUS: risk estimates and estimates of $\mfk{T}$ for ASUS at $n=5000$. Here $n_k^{\star}=|\mc{I}_{k}^{\tau^\star}|$ and $n_k=|\widehat{\mc{I}}_{k}^{\tau}|$ for $k=1,2$.}
	\scalebox{0.8}{
		\begin{tabular}{clcc}
			\toprule
			&       & \multicolumn{2}{c}{Asymptotic performance of ASUS} \\
			\midrule
			&       & Scenario S1    & Scenario S2 \\
			\midrule
			\multirow{4}[0]{*}{OR} & $\tau^\star$     & 2.00     & 0.980 \\
			& $t_1^\star$, $t_2^\star$ & 4.115, 0.130 & 4.073, 0.062 \\
			& $n_1^\star$ ,$n_2^\star$ & 4750, 250 & 4000, 1000 \\
			& risk  & 0.107 & 0.154 \\
			\midrule
			\multirow{4}[0]{*}{ASUS.1} & $\tau$   & 1.936 & 1.904 \\
			& $t_1$, $t_2$ & 1.253, 3.520 &  0.740, 1.281 \\
			& $n_1$, $n_2$ &  3460, 1540 & 2857, 2143  \\
			& risk  & 0.183 & 0.241 \\
			\midrule
			\multirow{4}[0]{*}{ASUS.2} & $\tau$   & 1.60 & 2.918 \\
			& $t_1$, $t_2$ & 0.420, 4.104 & 0.139, 1.8 \\
			& $n_1$, $n_2$ &   446, 4554   & 1024, 3976  \\
			& risk  & 0.126 & 0.161 \\
			\midrule
			\multirow{4}[0]{*}{Aux-Scr.1} & $\tau$   & 5.920 & 34.218 \\
			& $t_1$, $t_2$ & 0.424, - & 0,-  \\
			& $n_1$, $n_2$ &  5000, 0 & 5000,0  \\
			& risk  & 0.189 & 0.254 \\
			\midrule
			\multirow{4}[0]{*}{Aux-Scr.2} & $\tau$   & 7.375 & 50.958 \\
			& $t_1$, $t_2$ & 0.424, - &  0,- \\
			& $n_1$, $n_2$ &  5000, 0   & 5000, 0  \\
			& risk  & 0.189 & 0.254 \\
			\midrule
			SureShrink & risk  & 0.189 & 0.254 \\
			EBT   & risk  & 0.257 & 0.402 \\
			EJS   & risk  & 0.411 & 0.431 \\
			\bottomrule
	\end{tabular}}
	\label{tab:simexp3}%
\end{table}%
across both the scenarios, the risk of the auxiliary screening procedure, Aux-Scr, is indistinguishable from the risk of the SureShrink estimator, thus demonstrating that discarding observations based on the magnitude of the side information may lead to missing important signal coordinates especially when the side information is corrupted with noise. ASUS, however, does not discard any observations and continues to exploit the available information in the noisy auxiliary sequence. In table \ref{tab:simexp3}, we report risk estimates and estimates of $\mathcal{T}$ for ASUS and Aux-Scr at $n = 5000$. 

In table \ref{tab:simexp1_mainpaper}, we report the estimates of the hyper-parameters of Aux-Scr under the setting of the simulation experiment described in section 4.2.1 of the main paper.
\begin{table}[!t]
	\centering
	\caption{Risk estimates and estimates of $\mfk{T}$ for Aux-Scr at $n=5000$ under the setting of the simulation experiment described in section 4.2.1 of the main paper. Here $n_k^{\star}=|\mc{I}_{k}^{\tau^\star}|$ and $n_k=|\widehat{\mc{I}}_{k}^{\tau}|$ for $k=1,2$.}
	\scalebox{0.8}{
		\begin{tabular}{clcc}
			\toprule
			&       & \multicolumn{2}{c}{Asymptotic performance of Aux-Scr} \\
			\midrule
			&       & Scenario S1    & Scenario S2 \\
			\midrule
			\multirow{4}[0]{*}{Aux-Scr.1} & $\tau$   & 1.326 & 0.981 \\
			& $t_1$, $t_2$ & 4.612, 0.109 &  4.618, 0.159 \\
			& $n_1$, $n_2$ &  4747, 253 & 4002, 998  \\
			& risk  & 0.097 & 0.243 \\
			\midrule
			\multirow{4}[0]{*}{Aux-Scr.2} & $\tau$   & 11.231 & 10.752 \\
			& $t_1$, $t_2$ & 4.601, 0.106 & 4.580, 0.160 \\
			& $n_1$, $n_2$ &   4748, 252   & 3976, 1024  \\
			& risk  & 0.095 & 0.232 \\
			\midrule
			\multirow{4}[0]{*}{Aux-Scr.3} & $\tau$   & 1.774 & 1.460 \\
			& $t_1$, $t_2$ & 4.668, 0.665 & 4.760, 0.544  \\
			& $n_1$, $n_2$ &  4261, 739 & 3290, 1710  \\
			& risk  & 0.147 & 0.360 \\
			\midrule
			\multirow{4}[0]{*}{Aux-Scr.4} & $\tau$   & 5.072 & 2.004 \\
			& $t_1$, $t_2$ & 3.036, 1.043 &  3.500, 0.851 \\
			& $n_1$, $n_2$ &  2023, 2977   & 984, 4016  \\
			& risk  & 0.186 & 0.414 \\
			\bottomrule
	\end{tabular}}
	\label{tab:simexp1_mainpaper}%
\end{table}%
\section{Microarray Time Course (MTC) Data}
\label{supp:mtc}
Our second real data application is an MTC dataset collected by \citet{calvano2005network} for studying systemic inflammation in humans and is an example of a setting where ASUS can be used for 2 sample estimation problems. This dataset contains eight study subjects which are randomly assigned to case and control groups and then administered with endotoxin and placebo, respectively. The expression levels of $n=22,283$ genes in human leukocytes are measured before infusion (0 hour) and at 2, 4, 6, 9, and 24 hours afterwards. One of the goals of this experiment is to identify, in the case group, early to middle response genes that are differentially expressed within 4 hours and thus reveal meaningful early activation gene sequence that governs the immune responses. As discussed in \citet{sun2011multiple} the early activation sequence quickly activates many secreted pro-inflammatory factors in response to exterior intrusion. These activated factors subsequently trigger the expression of several transcription factors to initiate the immune response. In the late period, the expression levels of a number of transcription factors limiting the immune response are increased. Finally the whole system concludes with full recovery and a normal phenotype.
\begin{figure}
	\centering
	\begin{subfigure}[b]{\textwidth}
		\includegraphics[width=0.9\linewidth]{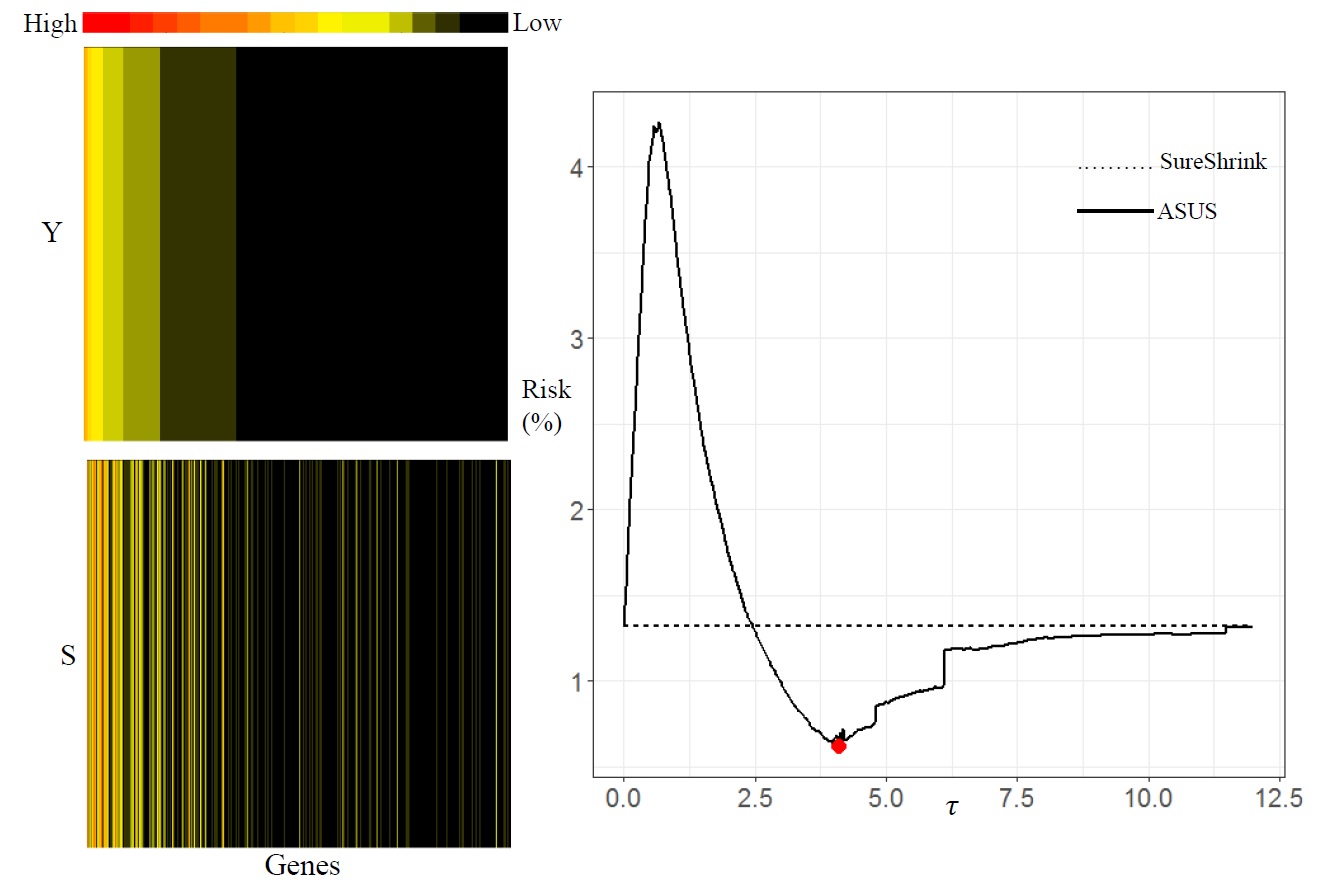}
		\caption{}
		\label{mtc1} 
	\end{subfigure}
	\begin{subfigure}[b]{\textwidth}
		\includegraphics[width=0.9\linewidth]{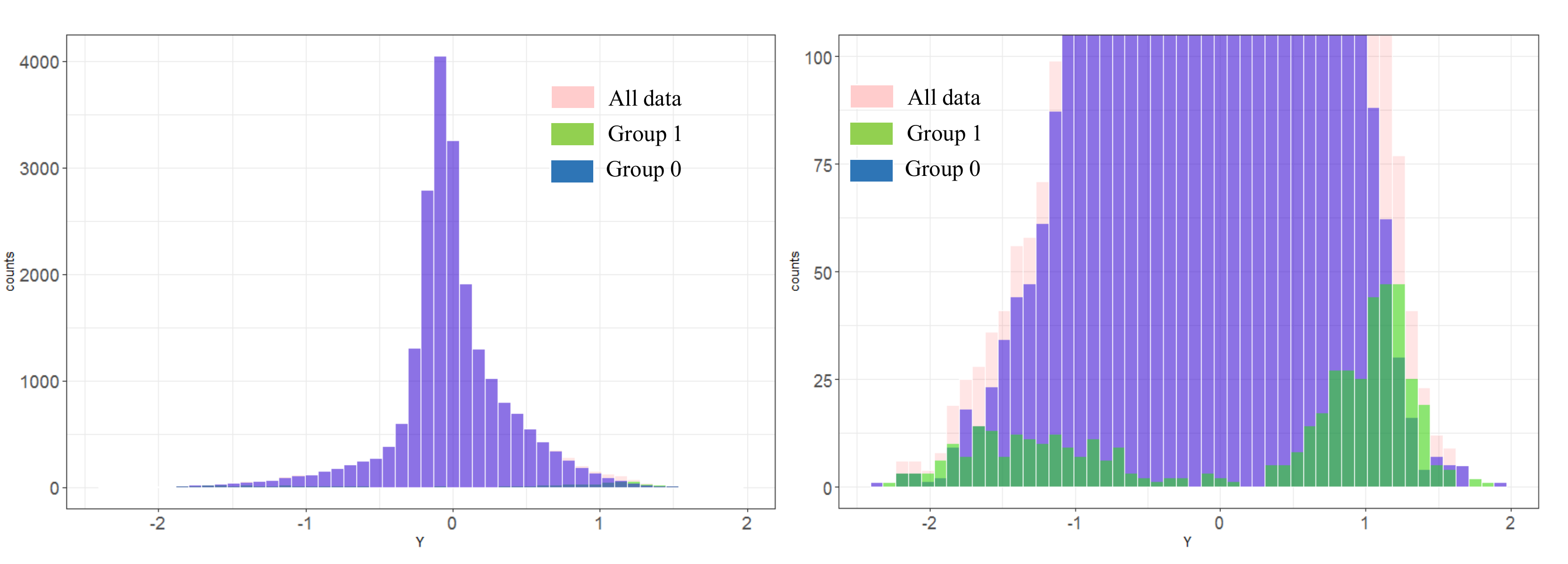}
		\caption{}
		\label{mtc2}
	\end{subfigure}
	\caption[Two numerical solutions]{(a) Left: Heatmap of gene expressions $\pmb Y$ and side information $\bm S$. Right: SURE estimate of the risk of $\hat{\theta}^{S}_i(t)$ at $t=1.13$ versus an unbiased estimate of the risk of ASUS for different values of $\tau$. (b) Left: Histogram of gene expressions $\pmb Y$. Group 1 is $\widehat{\mc{I}}^{\tau}_2$ and Group 0 is $\widehat{\mc{I}}^{\tau}_1$. Right: A magnified plot to show $\widehat{\mc{I}}^{\tau}_2$.}
\end{figure}
To identify the genes that regulate this sequence, we take time point 0 as the baseline and time points 4 and 24 as the interval over which differential gene expression will be estimated. We follow the data preprocessing steps outlined in \cite{sun2011multiple} and denote $Y_{i,j}$ as the arcsinh transformed average gene expression value for gene $i$ at time point $j$. Let $Y_i=\tilde{Y}_{i,4}-\tilde{Y}_{i,24}$ where $\tilde{Y}_{i,j}=Y_{i,j}-Y_{i,0}$ denotes the baseline adjusted expression level of gene $i$ at time point $j$. The side information that we use in this setting is $X_{i}=\tilde{Y}_{i,4}+\kappa_i\tilde{Y}_{i,24}$ with ${S}_i=|X_i|$, $K=2$, $\kappa_i=\tilde{\sigma}_{i,4}/\tilde{\sigma}_{i,24}$ where $\tilde{\sigma}_{i,j}$ is the observed standard deviation of $\tilde{Y}_{i,j}$ across the 4 replicates. In figure \ref{mtc1} right, the dotted line represents the SURE estimate of risk of $\hat{\bm{\theta}}^S(t)$ at $t=1.13$. ASUS uses information in $(Y_i,{S}_i)$ and returns an estimate of risk (the red dot) that is significantly smaller than the risk estimate returned by $\hat{\bm{\theta}}^S(t)$. In order to evaluate the results
in a predictive framework, we next used 3 out of the 4 replicates for calibrating the hyper-parameters
and calculated the prediction errors of the ASUS and SureShrink procedures based on the held
out fourth replicate. Here, the risk reduction by ASUS compared to SureShrink is almost $9\%$

Figure \ref{mtc1} left presents heatmap of expression level $Y_i$ in the top panel and heatmap of the associated side information ${S}_i$ in the bottom panel for gene $i$. The expression levels for the genes are ordered in terms of their magnitude. Notice how the magnitude of side information in ${S}_i$ follows the pattern of the expression levels $Y_i$, largely indicating that $Y_i$ is small whenever ${S}_i$ is small. ASUS exploits this extra information in $S_i$ and thus performs better than the SureShrink estimator that only relies on the information in $Y_i$.  Figure \ref{mtc2} left presents the distribution of gene expression for genes that belong to the groups $\widehat{\mc{I}}_1^\tau$ and $\widehat{\mc{I}}_2^\tau$. In this example, group $\widehat{\mc{I}}_2^\tau$ holds only about $2\%$ of the $n$ genes and is therefore inconspicuous in this plot. We present a magnified version of this plot in right that demonstrates in green the distribution of gene expression for genes that belong to $\widehat{\mc{I}}_2^\tau$ and summarize the results in table \ref{tab:mtc}.  
\begin{table}[!h]
	\centering
	\caption{Summary of the performance of SureShrink and ASUS on MTC data. Here $n_k=|\widehat{\mc{I}}_{k}^{\tau}|$ for $k=1,2$.}
	\scalebox{0.8}{\begin{tabular}{cccc}
			\toprule
			&       & \multicolumn{1}{c}{MTC} \\
			\midrule
			& $n$     &  22,283\\ 
			\midrule
			\multirow{2}[0]{*}{SureShrink} & $t$     & 1.13 \\
			& SURE estimate & 1.32 \\
			\midrule
			\multirow{6}[0]{*}{ASUS} & $\tau$   & 4.11  \\
			& $t_1$  &  1.3\\
			& $t_2$  &  0.04\\
			& $n_1$ & 21,791 \\
			& $n_2$  & 492 \\
			& SURE estimate & 0.62 \\
			\bottomrule
	\end{tabular}}%
	\label{tab:mtc}%
\end{table}%
In this real data example, a reduction in risk is possible because ASUS has efficiently exploited the sparsity
information encoded in $\bm S$. This can be seen, for example, from the stark contrast between
the magnitudes of thresholding hyper-parameters $t_1$ and $t_2$ in table \ref{tab:mtc}. Moreover, the risk of Aux-Scr for this example was seen to be no better than the SureShrink estimator and thus has been excluded from the results reported in table \ref{tab:mtc}.
\section{Choice of $K$}
\label{supp:chooseK}
\begin{figure}[!h]
	\centering
	\includegraphics[width=1\linewidth]{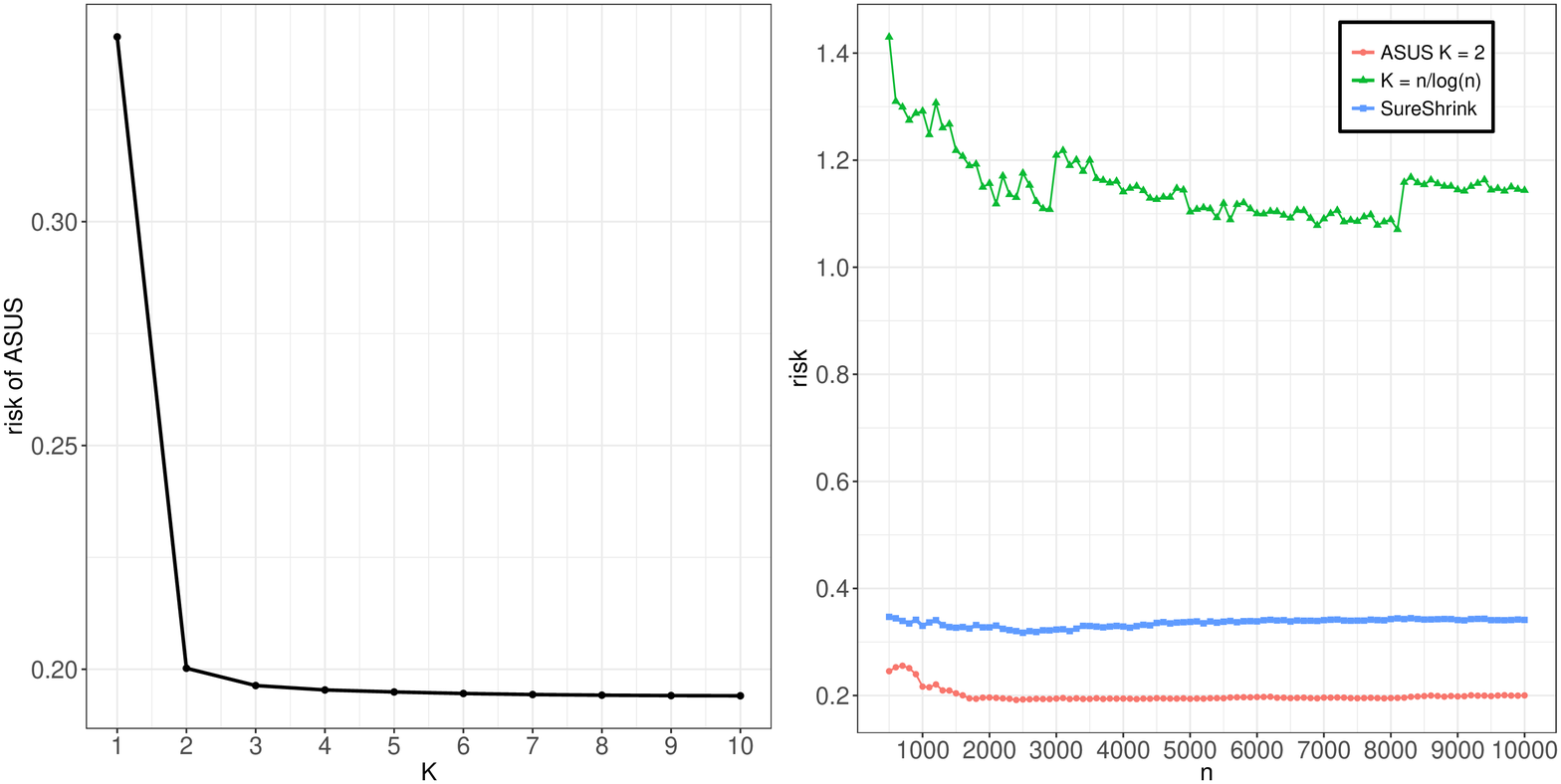}
	\caption{Toy example of section 2.3. Left: SURE estimate of risk of ASUS as $K$ varies. Right: Estimate of risks for ASUS with $K=2$, SureShrink and ASUS with no side information but with $K=n/\log n$.}
	\label{fig_K}
\end{figure}

We consider the toy example discussed in section 2.3 of the main paper and let $K\ge 2$. For each candidate value of $K$ we plot the SURE estimate of risk of ASUS in figure \ref{fig_K} left. An estimate of $K$ may be taken to be the one that appears at the elbow of this plot and that implies $\widehat{K}=2$. Often a large value of $K$, say $K=5$ or $6$, may continue to provide a marginal reduction in overall risk as opposed to $K=2$ as seen in this example but such a reduction in risk comes at a cost of increased computational burden of conducting a search over $O(m_n^{K-1})$ points for large $n$. A cross validation based approach for selecting $K$ in such scenarios is often useful. 
On a related note, in figure \ref{fig_K} right we demonstrate how ASUS reaps the benefits of adapting to the informativeness of side information. We continue with the toy example of section 2.3 and estimate the hyper-parameter $\mfk{T}$ for $K=2$. In contrast to this scheme, we also construct a version of ASUS where the number of groups $K$ is automatically set to $n/\log n$ without the aid of any side information and only the thresholding hyper-parameters $t_k$ are determined using the data driven hybrid scheme of \cite{donoho1995adapting}. The risk of this estimator is denoted by the green dotted line in figure \ref{fig_K} which clearly indicates that ASUS provides a better risk performance when the segmentation hyper-parameter $\bm{\tau}$ is chosen in a data driven adaptive fashion.  

\bibliographystyle{chicago}
\bibliography{paperref-supp}